\documentclass[11pt]{article}

\usepackage[T1,T2A]{fontenc}
\usepackage[cp1251]{inputenc}
\usepackage{textcomp}
\usepackage[centertags]{amsmath}
\usepackage[mediummath]{nccmath}
\usepackage{amsfonts}
\usepackage{amssymb}
\usepackage{braket}
\usepackage[pdftex]{hyperref}
\usepackage{graphicx}
\usepackage{graphbox}
\usepackage[numbers,sort&compress]{natbib}

\usepackage{paperinitial}

\paperinitialization{15mm}{15mm}{15mm}{15mm}{2pt}{10pt}

\DeclareMathOperator{\re}{Re}
\DeclareMathOperator{\im}{Im}

\DeclareMathOperator{\Li}{Li}
\DeclareMathOperator{\erfc}{erfc}

\newcommand{\e}{\varepsilon}

\newcommand{\s}{\sigma}

\newcommand{\al}{\alpha}
\newcommand{\be}{\beta}

\newcommand{\Ga}{\Gamma}
\newcommand{\de}{\delta}

\newcommand{\vk}{\varkappa}
\newcommand{\la}{\lambda}

\newcommand{\ups}{\upsilon}
\newcommand{\tmu}{\tilde{\mu}}

\newcommand{\spx}{\mathbf{x}}
\newcommand{\spy}{\mathbf{y}}

\newcommand{\spp}{\mathbf{p}}
\newcommand{\spq}{\mathbf{q}}
\newcommand{\spk}{\mathbf{k}}

\def\Xint#1{\mathchoice
{\XXint\displaystyle\textstyle{#1}}%
{\XXint\textstyle\scriptstyle{#1}}%
{\XXint\scriptstyle\scriptscriptstyle{#1}}%
{\XXint\scriptscriptstyle\scriptscriptstyle{#1}}%
\!\int}
\def\XXint#1#2#3{{\setbox0=\hbox{$#1{#2#3}{\int}$}
\vcenter{\hbox{$#2#3$}}\kern-.5\wd0}}

\def\dashint{\Xint-}

\begin{document}
\allowdisplaybreaks[4]
\frenchspacing
\setlength{\unitlength}{1pt}


\title{{\Large\textbf{Plasmon-polaritons in a rarefied nonrelativistic neutron gas}}}

\date{}

\author{%
P.O. Kazinski\thanks{E-mail: \texttt{kpo@phys.tsu.ru}}\;
and
A.M. Trushchuk\thanks{E-mail: \texttt{158erti08@gmail.com}}\\[0.5em]
{\normalsize Physics Faculty, Tomsk State University, Tomsk 634050, Russia}
}

\maketitle

\begin{abstract}

The electromagnetic properties of a dilute unpolarized nonrelativistic neutron gas are described in the case when neutron-by-neutron scattering is negligible. The Gaussian (Maxwell-Boltzmann) and Fermi-Dirac one-particle density matrices for the neutron gas are considered, the typical space scale of variations of these density matrices being assumed to be much larger than the typical space scale of variations of the electromagnetic field. The Green functions for the static effective Maxwell equations are obtained. It is shown that a rarefied neutron gas does not possess a ferromagnetic instability even in a nonequilibrium state. The analog of Friedel oscillations is described. The properties of plasmon-polaritons in the neutron gas are revealed. It turns out that the dispersion law of longitudinal and transverse plasmon-polaritons has an infinite number of branches at a given momentum. The region of parameters where the plasmon-polaritons can be regarded as quasiparticles is found. The plasmon-polaritons on a Gaussian wave packet of a single electron are described. Their dispersion law proves to have an infinite number of branches at a given momentum. It is shown that there exist stable longitudinal plasmon-polaritons even on a single electron.

\end{abstract}

\section{Introduction}

At present, there are the developed techniques allowing to produce the beams of neutrons of a rather high density with various neutron energies, from ultracold with kinetic energies less than $3\times10^{-7}\, \text{eV} \approx 3.5\,  \text{mK}$ up to ultrarelativistic, that are used in diverse applications \cite{Ignatovich1996,Serebrov2011,Pokotilovski2018,Lauss2021,Henderson2014,Anderson2016}. Furthermore, a neutron gas with the density much lower than the nuclear one arises in the neighborhood of neutron stars and in the processes at early stages of their formation \cite{HaensPtoYak2007,Potekhin2010,Chatziioannou2025}. The electromagnetic properties of a neutron gas  with such densities appear to have not been studied before excepting the magnetic susceptibility at zero momentum for a neutron gas in a thermodynamic equilibrium \cite{Clark1969,Delsante1979,Anand1981}. In describing the electromagnetic properties of a neutron gas near neutron stars, it is commonly supposed that its density is of order of a nuclear one and so the main effect on these properties is caused by the nuclear interaction between neutrons \cite{Chatziioannou2025,HaensPtoYak2007,Potekhin2010,Broderick2000,Kapusta2006book,Baldo2009,Gezerlis2010,Dong2013,Gandolfi2015,Vidana2021,Akhiezer1996}. Along with this, as a rule, the influence of other particles (the protons, electrons, muons, etc.) and the electroweak processes driving the system to a thermodynamic equilibrium are taken into account. In the present paper, we focus on the study of the other parameter domain of a neutron gas where its density is small in comparison with the nuclear one and the collisions between neutrons can be neglected. We will refer to such a neutron gas as rarefied or dilute and will consider the gas consisting only of neutrons at the time scales when the beta decay is inessential.

Despite the fact that neutrons do not have an electric charge, they interact with the electromagnetic field by means of the presence of their anomalous magnetic moments. As we shall see, this leads to a nontrivial response of a neutron gas to the external electromagnetic field, in many ways similar to the response of an electron plasma, even in the case when this gas is dilute. In particular, just as in an electron plasma, there are the plasmon-polaritons in a rarefied neutron gas that are the perturbations of the electromagnetic field with the dispersion law substantially deviating from the vacuum one. The electromagnetic properties of a neutron gas are completely described by the photon polarization operator. The general expression for it was derived in \cite{ComptNeutr} and we shall use it in the present paper. Notice that neutrons were described in \cite{ComptNeutr} as point Dirac particles with anomalous magnetic moment. Such an approximation is widely used in describing the electromagnetic interaction of neutrons at small momentum transfer to a photon in comparison with the pi-meson mass \cite{Dong2013,Broderick2000,Clark1969,Delsante1979,Anand1981} and it follows from the soft theorems in quantum field theory \cite{WeinbergB.12,Low1954,GellMann1954,Weinberg1970}.

As in an electron plasma, the presence of plasmon-polaritons in a neutron gas is associated with the presence of singularities in the photon-neutron scattering amplitude out of the photon mass-shell. Because of this, the photon polarization operator possesses singularities in the case of a narrow in momentum space quantum state of neutrons that indicate the presence of quasiparticles in the theory (see, e.g., \cite{WeinbergB.12}). By analogy with the electron plasma, these quasiparticles can be called plasmons \cite{ComptNeutr} and their hybridization with the electromagnetic field gives rise to the existence of plasmon-polaritons. We shall see below that these plasmon-polaritons with relatively large lifetime have the dispersion law close to the dispersion law of plasmons. Moreover, we shall see that such plasmon-polaritons are present even on a wave packet of a single neutron. The formalism developed in \cite{AKS2025,ComptNeutr} is applicable to describing the electromagnetic properties of a neutron gas prepared in an arbitrary quantum state, including the one-particle one, that allows for the standard perturbation theory in the coupling constant.

To demonstrate the similarities and differences of the properties of plasmon-polaritons in a neutron gas with the properties of plasmon-polaritons in an electron plasma, we also consider some properties of plasmon-polaritons on a Gaussian wave packet of a single electron. The general expression for the photon polarization operator in the presence of an electron gas prepared in an arbitrary quantum state inhomogeneous in the coordinate space and polarized with respect to spin has been obtained apparently for the first time in \cite{AKS2025}. In the case of an electron state homogeneous in space and unpolarized with respect to spin, it coincides with the expressions known in the literature \cite{Lindhard1954,Silin1960,Tsytovich1961,Braaten1993,Melrose2008,VladTysh2011}. Employing the expression for the photon polarization operator in the presence of a single electron obtained in \cite{AKS2025}, we show that there are the plasmon-polaritons on a wave packet of a single electron with an infinite lifetime within the approximations we use.

Besides the plasmon-polaritons in a neutron gas, we investigate the static limit of the effective Maxwell equations in detail. We obtain the explicit expressions for the magnetic and electric susceptibilities, with account for the spatial dispersion, for a neutron gas obeying the Maxwell-Boltzmann or Fermi-Dirac distributions. In the particular case of a degenerate neutron gas, the expression for magnetic susceptibility at zero momentum coincides with the known one in the literature \cite{Clark1969}. For a neutron gas with the Maxwell-Boltzmann distribution, the expression for magnetic susceptibility agrees by the order of magnitude with the expression presented in \cite{Anand1981}, where the explicit expression for magnetic susceptibility at zero momentum was derived for a relativistic neutron gas in the high-temperature limit $T\gg Mc^2$, i.e., in the parameter region different from that we investigate. Furthermore, in the nonrelativistic limit and at small plasmon-polariton energies, $|k_0|\ll|\spk|$, the transverse dielectric permittivity that we investigate coincides with the paramagnetic contribution to the transverse dielectric permittivity of a nonrelativistic electron gas obtained in \cite{Bobrov1992, Bobrov2018}.

The paper is organized as follows. In Sec. \ref{Pol_Oper_LDNG}, we provide the general expression for the photon polarization operator in the presence of a neutron gas unpolarized with respect to spin and discuss the one-particle density matrices of this gas, for which the photon polarization operator is investigated, and the approximations we use. Furthermore, we give here the general expressions for the effective Maxwell equations and the longitudinal and transverse parts of the polarization operator. In Sec. \ref{Pol_Oper_Gauss}, we derive the explicit expression for the photon polarization operator in the presence of a nonrelativistic neutron gas prepared in the quantum state with Gaussian one-particle density matrix. Section \ref{Stat_Lim_Gauss} is devoted to the static limit, whereas Secs. \ref{Long_Plams_Polar} and \ref{Trans_Plasm_Polar} are dedicated to the properties of plasmon-polaritons. In Sec. \ref{Neutrons_FD_Distr}, we obtain the explicit expression for the photon polarization operator for a neutron gas with the one-particle density matrix being the Fermi-Dirac distribution. In Sec. \ref{Stat_Lim_FD}, we consider the static limit while, in Sec. \ref{Plasm_Polar_FD}, we investigate the properties of plasmon-polaritons in the degenerate neutron gas. Some technicalities and the discussion of certain approximations used in the main text are moved to the appendices. Besides, in Appendix \ref{Plasm-Pol_on_Singl_El_App}, we discuss the properties of plasmon-polaritons on the Gaussian wave packet of a single electron and, in Appendix \ref{Asympt_k_to_0_App}, we obtain the long wavelength asymptotics of the polarization operator. We use the system of units such that $\hbar=c=1$ and $e^2=4\pi\al$, where $\al$ is the fine structure constant. The Minkowski metric, $\eta_{\mu\nu}$, is taken with the mostly minus signature.

\section{Polarization operator for a rarefied neutron gas}\label{Pol_Oper_LDNG}

In the paper \cite{ComptNeutr}, the expression for the photon polarization operator out of the photon mass-shell was derived in the presence of a low-density neutron gas, the photon momenta being assumed to be much less than the pi-meson mass, viz., $|k^\mu|\lesssim10$ MeV/c in the rest frame of a neutron gas. The derivation of the expression for this operator goes along the same lines as the derivation of the photon polarization operator in the presence of an electron gas in an arbitrary quantum state presented in \cite{AKS2025}. It is based on the application of the $in$-$in$ formalism \cite{Schw1961,Keld64,CSHY85,GFSh.3,DeWGAQFT.11,CalzHu} to a neutron gas prepared in an arbitrary quantum spate admitting the description in the framework of the standard perturbation theory. The photon polarization operator for a neutron gas is expressed through the time dependent one-particle density matrix of this gas in the momentum space $\rho_{ss'}(x^0;\spp,\spp')$. Let us introduce the Wigner function for this one-particle density matrix as
\begin{equation}
    \rho_{ss'}(x,\spp)=\int \frac{d\spq}{(2\pi)^3}e^{i\spq\spx}\rho^{(1)}_{ss'}(x^0;\spp+\spq/2,\spp-\spq/2).
\end{equation}
Decompose it into the scalar and spin parts,
\begin{equation}
    \rho_{ss'}(x,\spp)=\frac12\rho(x,\spp)\big[\de_{ss'} +\xi_a(x,\spp)(\s_a)_{ss'}\big],
\end{equation}
where $\s_a$ are the $\s$-matrices and $\xi_a(x,\spp)$ characterizes the spin polarization of a neutron gas. The normalization condition reads
\begin{equation}
    \int d\spx d\spp \rho(x,\spp)=N_h,
\end{equation}
where $N_h$ is the average number of neutrons in the gas. In the present paper, we will consider the electromagnetic properties of a spin unpolarized neutron gas, i.e., we will assume that $\xi_a(x,\spp)$ is negligibly small. Then the Weyl symbol of the polarization operator can be cast into the form \cite{ComptNeutr}
\begin{equation}\label{polar_oper0}
	\Pi^{\mu\nu}(x,k)=-\mu_p^2 M\int \frac{d\spp_c d\spq d\spy}{(2\pi)^3} e^{i\spq(\spx-\spy)}
    \frac{\rho(x^0,\spy,\spp_c)G_3^{\mu\nu}(\spp_c+\spq/2,\spp_c-\spq/2) \big|_{k\rightarrow k_+}}{\sqrt{p_0(\spp_c+\spq/2)p_0(\spp_c-\spq/2)}} ,\qquad p_0=\sqrt{M^2+\spp^2},
\end{equation}
where $\mu_p$ is the neutron anomalous magnetic moment, $M$ is the neutron mass, $k_+^\mu=k^\mu+i0\de^\mu_0$, and the expression for $G^{\mu\nu}_3$ is presented in formulas (46), (50) of \cite{ComptNeutr}. Notice that expression \eqref{polar_oper0} holds even for a single-neutron quantum state with $N_h=1$.

In the present paper, we shall derive the approximate expressions for the polarization operator \eqref{polar_oper0} and investigate some properties of the solutions to the respective effective Maxwell equations in the nonrelativistic approximation supposing that $\rho(x,\spp)$ is concentrated in the momentum domain $|\spp|\ll M$ in the reference frame where the neutron gas is at rest on average. We restrict our consideration to the one-particle density matrices with
\begin{subequations}\label{Wign_func_scalar}
\begin{align}
    i)\;\rho(\spx,\spp)& =\frac{N_h}{(2\pi)^3 \s^3_x\s^3} e^{-\frac{\spx^2}{2\s_x^2} -\frac{\spp^2}{2\s^2}}= \frac{\rho(\spx)}{(2\pi\s^2)^{3/2}} e^{-\frac{\spp^2}{2\s^2}},\label{Wign_func_scalar_MB}\\
    ii)\;\rho(x,\spp)&=\frac{2}{(2\pi)^3} \frac{1}{e^{\be(\spp^2/(2M)-\mu)}+1}= \Big(\frac{\be}{2\pi M} \Big)^{3/2} \frac{-1}{\Li_{3/2}(-e^{\be\mu})}\frac{\rho(x)}{e^{\be(\spp^2/(2M)-\mu)}+1},\label{Wign_func_scalar_FD}
\end{align}
\end{subequations}
where $\Li_\nu(z)$ is the polylogarithm (see some of its properties in Appendix \ref{App_F_D_Expansions}), $\rho(x)$ is the particle number density, and
\begin{subequations}
\begin{align}
    i)\; \rho(\spx)&=\frac{N_h}{(2\pi\s_x^2)^{3/2}} e^{-\frac{\spx^2}{2\s_x^2}},\\
    ii)\; \rho(x)&=-\frac14\Big(\frac{2M}{\pi\beta}\Big)^{3/2} \Li_{3/2}(-e^{\be\mu})\label{dens_neutr_gas}.
\end{align}
\end{subequations}
The first scalar Wigner function \eqref{Wign_func_scalar_MB} corresponds to a positive definite density matrix when $\s_x\s\geqslant1/2$ \cite{Bastos2012} and, as a particular case, describes the Maxwell-Boltzmann distribution with the reciprocal temperature $\be=M/\s^2$. This Wigner function evolves with time as
\begin{equation}
    \rho(x,\spp)=\rho\big(\spx -\frac{\spp}{M}x^0,\spp\big),\qquad \rho(x)=\frac{N_h}{\big[2\pi\s^2_x(x_0)\big]^{3/2}} e^{-\frac{\spx^2}{2\s_x^2(x_0)}},\qquad \s_x(x_0)=\s_x\sqrt{1+\Big(\frac{\s}{M}\frac{x_0}{\s_x}\Big)^2}.
\end{equation}
Henceforth, we will consider the evolution of the electromagnetic fields on the time scales where the dynamics of $\rho(x,\spp)$ can be neglected, viz., we will assume that
\begin{equation}
    \frac{\s}{M}|x^0|\ll\s_x.
\end{equation}
The second scalar Wigner function \eqref{Wign_func_scalar_FD} corresponds to the Fermi-Dirac distribution two-fold degenerate with respect to spin. It is stationary. This Wigner function describes the one-particle density matrix of a nonrelativistic gas of free neutrons in a thermodynamic equilibrium. For $\be\mu\ll-1$, i.e., for a sufficiently small density of the neutron gas, expression \eqref{Wign_func_scalar_FD} turns onto \eqref{Wign_func_scalar_MB}, where one should put $\s=\sqrt{M/\beta}$ (see for details Appendix \ref{App_F_D_Expansions}).

We will study the behavior of the electromagnetic field on the scales much less than the typical scale of variations of the scalar part of the Wigner function, $\rho(x,\spp)$, as a function of the variable $\spx$. As for the Wigner function \eqref{Wign_func_scalar_MB}, this scale of variations equals $\s_x$. The Wigner function of the Fermi-Dirac distribution \eqref{Wign_func_scalar_FD} possesses, formally, an infinite scale of variations with respect to $\spx$. Therefore, in this case we set $\s_x$ to the typical size of the domain where a thermodynamic equilibrium is reached. Then we have the restrictions
\begin{equation}\label{short_wave_appr}
    |\spk|\s_x\gg1,\qquad |k_0|\gg \frac{\s}{M\s_x},
\end{equation}
where one should substitute,
\begin{equation}\label{sigma_subs_FD}
    \s\rightarrow\max(\sqrt{M/\beta},p_F),\qquad p_F:=\sqrt{2M\mu},
\end{equation}
for the Fermi-Dirac distribution. The Fermi momentum, $p_F$, is defined when $\mu\geqslant0$. Moreover, there are the conditions
\begin{equation}\label{k_nonrel}
    \s\ll M,\qquad |\spk|\ll M,\qquad |k_0|\ll M,
\end{equation}
as long as we consider the nonrelativistic approximation.

As is seen from expression \eqref{polar_oper0} for the polarization operator, the main contribution to the integral comes from the region where
\begin{equation}
    |\spp_c|\lesssim \s,\qquad|\spq|\lesssim1/\s_x.
\end{equation}
The dependence of $G_3^{\mu\nu}$ in the numerator and of $p_0$ in the denominator of \eqref{polar_oper0} on $\spq$ can be discarded provided that
\begin{equation}\label{small_q_conds1}
    |q_0|\ll k_0,\qquad |q_0|\ll p^c_0,\qquad |\spq|\ll|\spk|,\qquad |\spq|\ll |\spp_c|.
\end{equation}
Since in the nonrelativistic approximation
\begin{equation}
    q_0=p_0(\spp_c+\spq/2) -p_0(\spp_c-\spq/2) \approx(\spp_c\spq)/M,
\end{equation}
the conditions \eqref{small_q_conds1} are satisfied if the estimates \eqref{short_wave_appr} hold and
\begin{equation}\label{small_q_sigma_x}
    \frac{\s}{M^2\s_x}\ll1,\qquad \s\s_x\gg1.
\end{equation}
The last estimate follows from the last condition in \eqref{small_q_conds1}.

This estimate is too strict for a pure neutron state \eqref{Wign_func_scalar_MB}, where $\s_x\s=1/2$ and the main contribution to the integral in the polarization operator gives the domain
\begin{equation}\label{q_less_sigma}
    |\spq|\lesssim\s.
\end{equation}
Nevertheless, it turns out that the dependence of $G_3^{\mu\nu}$ in the numerator and of $p_0$ in the denominator of \eqref{polar_oper0} on $\spq$ can also be neglected provided one requires the fulfillment of the estimates
\begin{equation}\label{large_k_conds}
    |\spk|\gg\s,\qquad|k_0|\gg\frac{\s^2}{M},
\end{equation}
instead of conditions \eqref{short_wave_appr}. The justification of this approximation is given in Appendix \ref{Small_q_Approx_App}.

As a result, under the fulfillment of conditions \eqref{short_wave_appr}, \eqref{k_nonrel}, \eqref{small_q_sigma_x}, or conditions \eqref{k_nonrel}, \eqref{q_less_sigma}, \eqref{large_k_conds}, the Weyl symbol of the polarization operator \eqref{polar_oper0} can be written with the help of formula (162) of \cite{ComptNeutr} as
\begin{equation}\label{polar_oper}
	\Pi^{\mu\nu}(x,k)=-\mu_p^2 M\int \frac{d\spp}{p_0}  \rho(x,\spp) G_3^{\mu\nu}(\spp,\spp)|_{k\rightarrow k_+},
\end{equation}
where, for brevity, we change the notation of the integration variable, $\spp_c\rightarrow\spp$, and
\begin{equation}\label{G3_off-shell}
	G_3^{\mu\nu}(\spp,\spp) =-\frac{M k^4}{(kp)^2-k^4/4} \Big[\eta^{\mu\nu}
	-\frac{p^\mu p^\nu}{M^2} +\frac{k^{(\mu}p^{\nu)}(kp)}{M^2k^2}
    -\frac{k^\mu k^\nu}{k^2}\Big(1+\frac{(kp)^2}{M^2k^2}\Big)\Big],
\end{equation}
where $k^4\equiv (k^2)^2=(k_\nu k^\nu)^2$. Hereinafter, the parentheses at a pair of indices mean a symmetrization without the factor $1/2$. The effective Maxwell equations take the form
\begin{equation}\label{Max_eq_eff_0}
    \big[\eta^{\mu\nu}\Box-\partial^\mu\partial^\nu +\Pi^{\mu\nu}(x,i \partial_x)\big]A_\nu(x)=0.
\end{equation}
The derivatives in the expression for $\Pi^{\mu\nu}(x,i \partial_x)$ ought to be Weyl ordered. However, when the conditions \eqref{short_wave_appr} or \eqref{large_k_conds} are satisfied, noncommutativity of $x$ and $\partial_x$ in $\Pi^{\mu\nu}(x,i \partial_x)$ entering Eq. \eqref{Max_eq_eff_0} can be neglected, and one can place all the derivatives $i\partial_x$ on the right of all $x$.

The expression \eqref{polar_oper} contains only the contribution to the polarization operator that depends on the neutron density matrix. As for the vacuum term, the main contribution to it stems from the one-loop electron-positron correction. This correction can be taken into account in the effective Maxwell equations by replacing $\mu_p^2$ by
\begin{equation}
    \mu^2_{eff}(k^2):=\mu_p^2/(1-\Pi(k^2_+)),
\end{equation}
where $\Pi(k^2)$ is the scalar part of the vacuum polarization operator (see formulas (24), (54) in \cite{AKS2025}). A more detailed exposition of the procedure how to take into account the vacuum contribution to the polarization operator can be found in \cite{AKS2025}, where the analogous procedure was carried out for the photon polarization operator in the presence of an electron gas. The correction appearing due to redefinition of $\mu_p^2$ is small for the photon momenta we consider and we will neglect it henceforth.

For the isotropic in the  momentum space scalar part of the Wigner function $\rho(x,\spp)$, the polarization operator \eqref{polar_oper} can be cast into the form (see, e.g., \cite{Kapusta2006book})
\begin{equation}
    \Pi^{\mu\nu}=\Pi_\parallel P^{\mu\nu}_\parallel +\Pi_\perp P^{\mu\nu}_\perp,
\end{equation}
where
\begin{equation}
\begin{gathered}
    P_\parallel^{\mu\nu}=\de_0^\mu \de_0^\nu -\frac{k^\mu k^\nu}{k^2} -\frac{(k^\mu-k_0\de_0^\mu) (k^\nu-k_0\de_0^\nu)}{\spk^2},\qquad P_\perp^{\mu\nu}=\eta^{\mu\nu} -\de_0^\mu \de_0^\nu +\frac{(k^\mu-k_0\de_0^\mu) (k^\nu-k_0\de_0^\nu)}{\spk^2},\\
    P_\parallel^{\mu\nu}k_\nu=P_\perp^{\mu\nu}k_\nu=0,\qquad P_\parallel^{\mu\nu}\ups_\nu=\ups^{\mu},\qquad P_\perp^{\mu\nu}\ups_\nu=0,\\
    P_\parallel^{\mu\nu}+P_\perp^{\mu\nu}=\eta^{\mu\nu}-\frac{k^\mu k^\nu}{k^2},\qquad  P^{\mu}_{\parallel\ \nu}P^{\nu}_{\perp\ \rho}= P^{\mu}_{\perp\ \nu}P^{\nu}_{\parallel\ \rho} =0,\qquad P_{\parallel\ \mu}^{\mu}=1,\qquad P_{\perp\ \mu}^{\mu}=2,
\end{gathered}
\end{equation}
and $\ups^\mu=\de_0^\nu-k^\mu k_0/k^2$. The projector $P_\parallel$ maps to the longitudinal plasmon-polariton modes, whereas $P_\perp$ projects to the transverse ones. Furthermore,
\begin{equation}
    \Pi_\parallel=-\frac{k^2}{\spk^2}\Pi^{00},\qquad \Pi_{\perp}=\frac12(\Pi^\mu_{\ \mu} -\Pi_\parallel).
\end{equation}
It follows from the general expression \eqref{polar_oper} that, in the isotropic case, we have for a dilute neutron gas unpolarized with respect to spin
\begin{equation}\label{Pi_par_Pi_perp_gen}
    \Pi_\parallel =\mu_p^2 k^2 \int\frac{d\spp}{p_0}\rho(x,\spp)\Big[1 - k^2 \frac{\spp_\perp^2 -k^2/4}{(kp)^2-k^4/4} \Big],\qquad \Pi_\perp =\mu_p^2M^2 k^4 \int\frac{d\spp}{p_0}\rho(x,\spp) \frac{1+\spp^2_\perp/(2 M^2)}{(kp)^2-k^4/4},
\end{equation}
where $\spp_\perp = \spp-(\spp\spk)\spk/\spk^2$ and $\im k_0=+0$. For the other values of $k_0\in \mathbb{C}$, the polarization operator is understood in the sense of an analytical continuation. Notice a high similarity of expressions \eqref{Pi_par_Pi_perp_gen} to the analogous quantities for an electron gas presented in Appendix \ref{Plasm-Pol_on_Singl_El_App}.

Performing Fourier transform of the effective Maxwell equations \eqref{Max_eq_eff_0}, these equation are reduced to the matrix equation under the assumptions discussed above,
\begin{equation}\label{Max_eq_eff}
    [-k^2\eta^{\mu\nu} +k^\mu k^\nu +\Pi^{\mu\nu}(k)] A_\nu(k)=0,
\end{equation}
where we have neglected the dependence of $\Pi^{\mu\nu}$ on $x$. This equation has the solution in the form of a longitudinal mode $\ups^\mu$ with the dispersion law
\begin{equation}\label{disp_law_longit0}
    \e_\parallel=0,
\end{equation}
where the longitudinal dielectric permittivity is defined as
\begin{equation}
    \e_\parallel=1-\frac{\Pi_\parallel}{k^2}.
\end{equation}
The transverse modes orthogonal to $\de_0^\mu$ and $k^\mu$ are also the solutions to \eqref{Max_eq_eff} and possess the dispersion law
\begin{equation}\label{disp_law_transv0}
    k^2-\Pi_\perp=0.
\end{equation}
Taking into account the explicit expression for $\Pi_\perp$ given in \eqref{Pi_par_Pi_perp_gen}, we see that there are the trivial transverse modes with the free dispersion law
\begin{equation}
    k^2=0,
\end{equation}
and the nontrivial transverse plasmon-polariton modes with the dispersion law
\begin{equation}\label{disp_law_transv}
    \mu^{-1}_\perp=0,
\end{equation}
where the magnetic permeability of transverse modes,
\begin{equation}\label{magn_permit}
    \mu_\perp=\Big(1-\frac{\Pi_\perp}{k^2}\Big)^{-1},
\end{equation}
has been introduced.

In the static limit, $k_0\rightarrow0$, the effective Maxwell equation for the electrostatic potential becomes
\begin{equation}\label{Max_eqs_stat_00}
    (\spk^2+\Pi^{00}(\spk))A^0(\spk)=\spk^2 \e_\parallel(\spk)A^0(\spk)=j^0(\spk),
\end{equation}
where $j^0(\spk)$ is the Fourier transform of the charge density. If the stationary current $j^i(\spk)$ is placed in the neutron gas and
\begin{equation}
    j^0=0,\qquad k_ij^i(\spk)=0,
\end{equation}
then, in the static limit and in the Coulomb gauge, $k_iA^i(\spk)$=0, the effective Maxwell equations are reduced to
\begin{equation}\label{Max_eq_stat_transv0}
    (\spk^2+ \Pi_\perp(\spk)) A^i(\spk)=\spk^2\mu^{-1}_\perp(\spk)=j^i(\spk).
\end{equation}
In the following sections, we shall thoroughly investigate the properties of solutions to the effective Maxwell equations \eqref{Max_eq_eff} for a neutron gas whose one-particle density matrices are described by the scalar Wigner functions \eqref{Wign_func_scalar}.

\section{Gaussian one-particle density matrix}\label{Pol_Oper_Gauss}

Let us obtain the explicit expression for the polarization operator \eqref{polar_oper} in the case of a Gaussian one-particle density matrix with scalar Wigner functions \eqref{Wign_func_scalar_MB} under the assumptions discussed in the previous section. Consider, at first, the transverse part of the polarization operator $\Pi_\perp$. If the first condition in \eqref{k_nonrel} holds, then $p_0$ can be replaced by $M$ in the denominator of the integrand of \eqref{Pi_par_Pi_perp_gen} and $\spp_\perp^2/(2M^2)$ in the numerator can be omitted. The integrand in \eqref{Pi_par_Pi_perp_gen} also contains
\begin{equation}
    \frac{1}{(kp)^2-k^4/4}=\frac{1}{k^2}\Big[\frac{1}{\al_- -\spk\spp+(p_0-M)k_0} -\frac{1}{\al_+ -\spk\spp+ (p_0-M)k_0}\Big],
\end{equation}
where
\begin{equation}
    \alpha_\pm=k_0M\pm k^2/2.
\end{equation}
We neglect the contribution of $(p_0-M)k_0$ in the denominator. This is justified if the following estimates are fulfilled,
\begin{equation}\label{nonrel_appr_2}
    |k_0|\ll\frac{M}{\s}|\spk|,\quad\text{or}\quad |k_0|\gtrsim \frac{M}{\s}|\spk|,\quad |k_0|\gg\frac{\s^2}{M},
\end{equation}
in addition to conditions \eqref{k_nonrel}. The conditions \eqref{nonrel_appr_2} ensure that the contribution of the term $(p_0-M)k_0$ is much less than the contributions of the retained terms $(\spk\spp)$ and $k^2/2$. Notice that, under the above approximations and at $|\spk_0|\ll|\spk|$, the expression for the transverse part of the polarization operator in \eqref{Pi_par_Pi_perp_gen} coincides with the paramagnetic contribution to the transverse part of the polarization operator of a nonrelativistic electron gas presented in \cite{Bobrov1992, Bobrov2018}.

In order to obtain the explicit expression for the polarization operator, we have to perform the integral
\begin{equation}\label{int_I}
	I(\alpha):=\int \frac{d\spp}{(2\pi)^{3/2}\s^3} \frac{e^{-\frac{\spp^2}{2\s^2}}}{\alpha-\spk \spp}, \qquad \im\alpha>0.
\end{equation}
Decomposing the vector $\spp$ into longitudinal and transverse parts with respect to the vector $\spk$ and integrating over the transverse components, we deduce
\begin{equation}\label{I_int}
	I(\alpha)=-\frac{1}{\alpha} F\Big(\frac{\alpha}{\sqrt{2}|\spk|\s}\Big),
\end{equation}
where (see, e.g., \cite{LandLifPhysKin,FriedConte1961})
\begin{equation}\label{plasm_funk}
	F(x)=\frac{x}{\sqrt{\pi}}\int_{-\infty}^{\infty}\frac{dz e^{-z^2}}{z-x},\quad\im x>0;\qquad
    F(x)=\frac{x}{\sqrt{\pi}}\int_{-\infty}^{\infty}\frac{dz e^{-z^2}}{z-x} +2\sqrt{\pi}ixe^{-x^2},\quad\im x<0.
\end{equation}
It is clear that this function is an entire function of $x$ and the symmetry properties hold
\begin{equation}\label{F_symm_rels}
    F(-x)=F(x) -2i\sqrt{\pi}xe^{-x^2},\qquad F^*(x)=F(x^*) -2i\sqrt{\pi}x^*e^{-(x^*)^2} =F(-x^*).
\end{equation}
The function \eqref{plasm_funk} is expressed in terms of the error function
\begin{equation}
    F(x)=i\sqrt{\pi}x e^{-x^2} (2-\erfc(ix)),
\end{equation}
where $\erfc(z)$ is the complementary error function. This expression is valid for any $x\in \mathbb{C}$.

To obtain the approximate expressions for the dispersion laws of plasmon-polaritons in a rarefied neutron gas, we shall need the asymptotic expansions
\begin{subequations}
\begin{align}
    F(x)&\simeq -\sum_{n=0}^\infty\frac{\Ga(1/2+n)}{\Ga(1/2)} x^{-2n}\approx -1-\frac{1}{2x^2},&\qquad|\arg(-ix)|&<\pi/4,\label{F_asympt}\\
    F(x)&\simeq -\sum_{n=0}^\infty\frac{\Ga(1/2+n)}{\Ga(1/2)} x^{-2n} +i\sqrt{\pi} xe^{-x^2} \approx -1-\frac{1}{2x^2} +i\sqrt{\pi} xe^{-x^2},&\qquad \re x^2&>0,\label{F_asympt_2}\\
    F(x)&\simeq -\sum_{n=0}^\infty\frac{\Ga(1/2+n)}{\Ga(1/2)} x^{-2n} +2i\sqrt{\pi} xe^{-x^2} \approx -1-\frac{1}{2x^2} +2i\sqrt{\pi} xe^{-x^2},&\qquad|\arg(ix)|&<\pi/4,\label{F_asympt_3}
\end{align}
\end{subequations}
where $|x|^2\gg1$. For a small argument,
\begin{equation}\label{F_small_arg}
	F(x)\approx i\sqrt{\pi} x -2x^2 -i\sqrt{\pi} x^3,\qquad |x|\ll1.
\end{equation}

Thus, under the above assumptions, the transverse part of the polarization operator reads
\begin{equation}\label{Pi_perp_Max}
    \Pi_\perp\approx k^2\Phi(k),
\end{equation}
where
\begin{equation}\label{Phi_k_Gauss}
    \Phi(k):=\mu^2_p(x)
    M\Big[ \frac{1}{\alpha_+} F\Big(\frac{\alpha_+}{\sqrt{2}|\spk|\s}\Big) -\frac{1}{\alpha_-}F\Big(\frac{\alpha_-}{\sqrt{2}|\spk|\s}\Big) \Big],
\end{equation}
and $\mu_p^2(x):=\mu_p^2\rho(x)$ is the density of the neutron anomalous magnetic moment squared. Notice that the equations,
\begin{equation}\label{plasm_disp_law}
    \alpha_{\pm}=0,
\end{equation}
describe the dispersion law of plasmons in the neutron gas \cite{ComptNeutr}
\begin{equation}\label{plasmon_disp_law}
    k_0=\pm\sqrt{M^2+\spk^2} \mp M\approx \pm\spk^2/(2 M),
\end{equation}
where the signs ``$\pm$'' are agreed with the signs in the subscript of $\al_\pm$ and only those solutions are kept that obey  condition \eqref{k_nonrel} on the photon momenta.

When the photon momenta satisfy the conditions \eqref{short_wave_appr} or \eqref{large_k_conds}, we can neglect the dependence of $\Pi^{\mu\nu}(x,k)$ on $x$ in the effective Maxwell equations, which on performing the Fourier transform turns into \eqref{Max_eq_eff}. The dispersion law of nontrivial transverse modes \eqref{disp_law_transv} is reduced to
\begin{equation}\label{transv_disp_law}
    \Phi(k)=1,
\end{equation}
and the magnetic permeability of transverse modes \eqref{magn_permit} is written as
\begin{equation}\label{mu_perp_Phi}
    \mu_\perp(k)=[1-\Phi(k)]^{-1}.
\end{equation}

Let us find the longitudinal part of the polarization operator $\Pi_\parallel$. Supposing that conditions \eqref{k_nonrel}, \eqref{nonrel_appr_2} hold, we have approximately
\begin{equation}
    \Pi_\parallel =\frac{\mu_p^2(x)}{M} k^2 \int \frac{d\spp}{(2\pi)^{3/2}\s^3} e^{-\frac{\spp^2}{2\s^2}} \Big[1 - (\spp_\perp^2 -k^2/4) \Big(\frac{1}{\al_- -\spk\spp} -\frac{1}{\al_+ -\spk\spp}\Big) \Big].
\end{equation}
As long as
\begin{equation}
    \int \frac{d\spp \spp_\perp^2}{(2\pi)^{3/2}\s^3} \frac{e^{-\frac{\spp^2}{2\s^2}}}{\alpha-\spk \spp} =2\s^2 I(\al),
\end{equation}
the longitudinal part of the polarization operator becomes
\begin{equation}\label{Pi_parallel_Max}
    \Pi_\parallel =-k^2\chi_\parallel(k),
\end{equation}
where the longitudinal electric susceptibility,
\begin{equation}\label{chi_parallel0}
    \chi_\parallel(k) \approx -\frac{\mu_p^2(x)}{M} +\frac{2\s^2}{M^2}\Big(1-\frac{k^2}{8\s^2}\Big)\Phi(k),
\end{equation}
has been introduced. The dispersion law of longitudinal modes \eqref{disp_law_longit0} turns into
\begin{equation}\label{longitud_disp_law}
    \chi_\parallel(k)=-1.
\end{equation}
One should discard the solutions of Eq. \eqref{longitud_disp_law} satisfying $k^2=0$, if they exist, since in this case the potential $\ups^\mu$ is a pure gauge.

The last symmetry property of the function $F(x)$ in \eqref{F_symm_rels} implies that, along with the solution $k_0(\spk)$ of Eqs. \eqref{transv_disp_law} or \eqref{longitud_disp_law} determining the dispersion law of plasmon-polaritons, these equations possess the solution $-k_0^*(\spk)$. Therefore, it is sufficient to consider the domain $\re k_0\geqslant0$ in investigating the dispersion laws.

\subsection{Static limit}\label{Stat_Lim_Gauss}

Before proceeding to the analysis of the dispersion laws of longitudinal and transverse plasmon-polaritons, we consider the static limit, $k_0\rightarrow0$, in Eqs. \eqref{Max_eq_eff}. The long wavelength asymptotics, $\spk\rightarrow0$, of the photon polarization operator in the presence of a uniform in space dilute neutron gas is given in Appendix \ref{Asympt_k_to_0_App}. In the static limit,
\begin{equation}\label{alpha_pm_stat}
    \al_\pm\approx \mp\spk^2/2,
\end{equation}
and
\begin{equation}\label{Phi_k_stat}
    \Phi(k)\approx -\frac{2\mu^2_p(x)M}{\spk^2} \Big[ F\Big(-\frac{|\spk|}{2\sqrt{2}\s}\Big) +F\Big(\frac{|\spk|}{2\sqrt{2}\s}\Big) \Big] =-\frac{\vk}{4} \frac{F(x)+F(-x)}{x^2}\Big|_{x=|\spk|/(2\sqrt{2}\s)},
\end{equation}
where
\begin{equation}
    \vk:=\Phi(0)=\frac{\mu_p^2(x)M}{\s^2}.
\end{equation}
The function $\Phi(k)$ defines the transverse magnetic permeability \eqref{mu_perp_Phi} and the longitudinal electric susceptibility \eqref{chi_parallel0}. For $|\spk|\gg \s$, we have
\begin{equation}\label{Phi_k_stat_large_arg}
    \Phi(k)\approx \frac{\vk}{2x^2}\Big|_{x=|\spk|/(2\sqrt{2}\s)},
\end{equation}
while for $|\spk|\ll\s$, we obtain
\begin{equation}\label{Phi_k_stat_small_arg}
    \Phi(k)\approx \vk.
\end{equation}
The static longitudinal electric susceptibility can be cast into the form
\begin{equation}\label{chi_par_stat}
    \chi_\parallel(\spk)=-\chi_\parallel(0) \Big[1+\frac{1+x^2}{2} \frac{F(x)+F(-x)}{x^2} \Big]_{x=|\spk|/(2\sqrt{2}\s)},
\end{equation}
where
\begin{equation}
    \chi_\parallel(0)=\frac{\mu_p^2(x)}{M}.
\end{equation}
For $|\spk|\gg \s$, we have
\begin{equation}\label{chi_stat_large_arg}
    \chi_\parallel(\spk) \approx \frac{3\chi_\parallel(0)}{2x^2}\Big|_{x=|\spk|/(2\sqrt{2}\s)},
\end{equation}
whereas for $|\spk|\ll\s$, we obtain
\begin{equation}\label{chi_stat_small_arg}
    \chi_\parallel(\spk) \approx \chi_\parallel(0)\Big(1+\frac{2x^2}{3} -\frac{4x^4}{5}\Big) _{x=|\spk|/(2\sqrt{2}\s)}.
\end{equation}
Notice that $\chi_\parallel(0)\ll \Phi(0)$.

We start with the description of a behavior of the transverse modes. The equation \eqref{Max_eq_stat_transv0} is written as
\begin{equation}\label{Max_eq_stat_transv}
    \spk^2\big[1-\Phi(\spk)\big] A^i(\spk)=j^i(\spk).
\end{equation}
This equation does not have solutions when $j^i(\spk)$ contains the modes with momenta $\spk$ satisfying the condition
\begin{equation}\label{instab_cond_stat}
    1-\Phi(\spk)=0.
\end{equation}
If $j^i(\spk)$ does not have the modes with such momenta, then the solution to Eq. \eqref{Max_eq_stat_transv} is ambiguous: one can always add $c^i(\spk)f(\spk)$ to the solution $A^i(\spk)$, where $c^i(\spk)$ is an arbitrary transverse vector and $f(\spk)$ is a generalized function concentrated on the set of points $\spk$ obeying \eqref{instab_cond_stat}. So long as $\Phi(\spk)=\Phi(|\spk|)$ in the case we consider, the momenta satisfying Eq. \eqref{instab_cond_stat} constitute a sphere.

Using the explicit expression \eqref{Phi_k_stat} for $\Phi(\spk)$, it is not difficult to ascertain that condition \eqref{instab_cond_stat} can be satisfied only if
\begin{equation}\label{instab_cond_stat_1}
    \Phi(0)=\vk\geqslant1.
\end{equation}
In this case, there is the only solution $|\spk|$ to Eq. \eqref{instab_cond_stat}. In the general case, $j^i(\spk)$ possesses the modes with momenta $\spk$ satisfying \eqref{instab_cond_stat} and the absence of static solutions to the effective Maxwell equations in this case indicates instability of a neutron gas when condition \eqref{instab_cond_stat_1} holds.

Let us find the Green function for Eq. \eqref{Max_eq_stat_transv} in the case $\vk\in(0,1)$. Taking into account the explicit expression for $\Phi(k)$ in the static limit \eqref{Phi_k_stat}, the Green function can be written as
\begin{equation}\label{Green_func_0}
    G(\spx)=\int\frac{d\spk}{(2\pi)^3} \frac{e^{i\spk\spx}}{\spk^2[1-\Phi(|\spk|)]}=\frac{1}{4\pi r} \frac{1}{1-\vk} +\frac{1}{4\pi^2ir} \int_{-\infty+i0}^{\infty+i0} \frac{dk}{k}\frac{e^{ikr}}{1-\Phi(k)},
\end{equation}
where $r=|\spx|$. Closing the integration contour in the second term in the upper half-plane, we obtain the expression for the Green function as a sum over the residues of the integrand. The equation \eqref{instab_cond_stat} determining the positions of poles has an infinite number of solution for $k\in \mathbb{C}$, $\im k>0$. For large $r$, the contribution of the second term in \eqref{Green_func_0} is exponentially suppressed and its leading asymptotics is specified by the residue of the function $1/(1-\Phi(k))$ closest to the real axis.


\begin{figure}[tp]
\centering
\includegraphics*[width=0.47\linewidth]{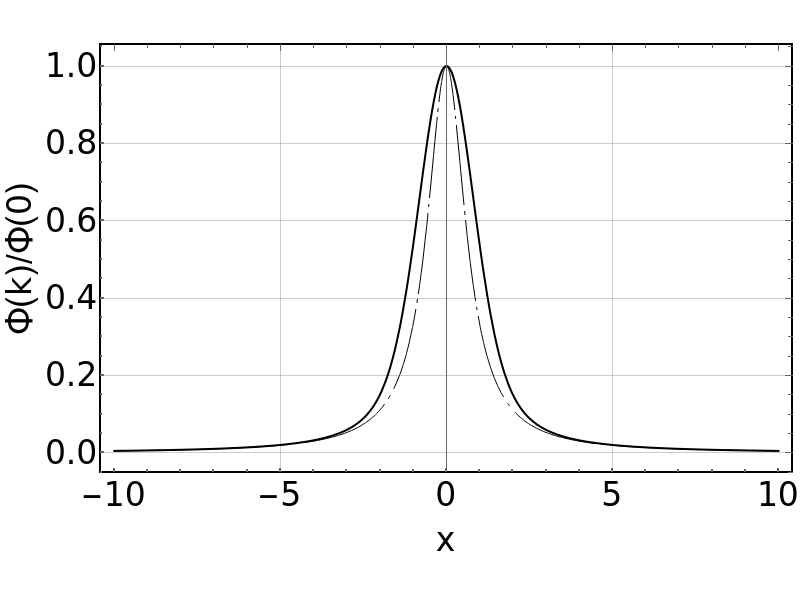}\;
\includegraphics*[width=0.47\linewidth]{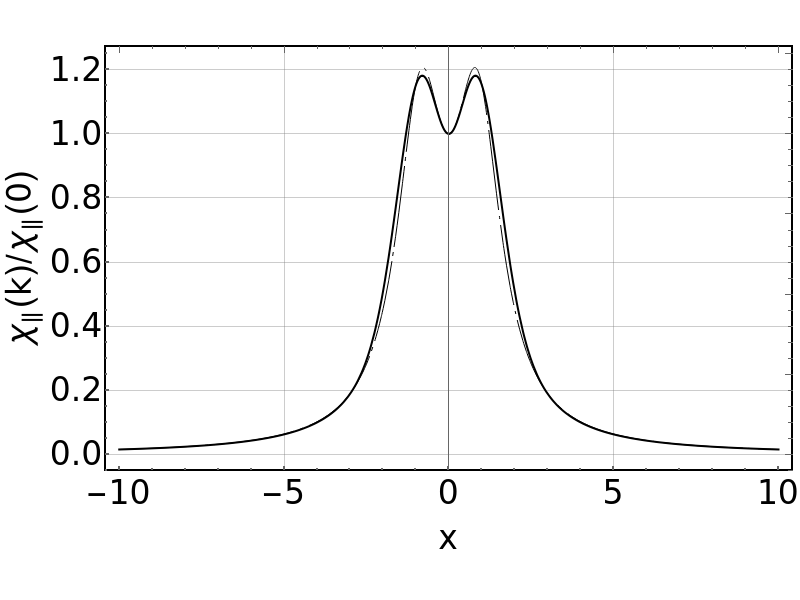}
\caption{{\footnotesize The functions $\Phi(k)$ and $\chi_\parallel(k)$ for the Gaussian one-particle density matrix of a neutron gas \eqref{Wign_func_scalar_MB} in the static limit. Here $x=k/(2\sqrt{2}\s)$. The solid lines are the exact values, whereas the dashed-dotted lines are the corresponding approximations \eqref{Phi_k_approx} and \eqref{chi_k_approx}.}}
\label{Phi_Chi_stat_MB_plots}
\end{figure}


In order to find the explicit approximate expression for the Green function, we employ the asymptotics \eqref{Phi_k_stat_large_arg}, \eqref{Phi_k_stat_small_arg} of the function $\Phi(k)$ for large and small arguments. Joining these asymptotics (see Fig. \ref{Phi_Chi_stat_MB_plots}), we have approximately
\begin{equation}\label{Phi_k_approx}
    \Phi(k)\approx \frac{\vk}{1+2 x^2}\Big|_{x=k/(2\sqrt{2}\s)}.
\end{equation}
Then the integral in \eqref{Green_func_0} can be easily performed and we arrive at
\begin{equation}\label{Green_func_transv0}
    G(\spx) \approx \frac{1}{4\pi r}\frac{1}{1-\vk} \big( 1-\vk e^{-2\sqrt{1-\vk}\s r}\big).
\end{equation}
As we see, for $\vk\in(0,1)$, a nonrelativistic neutron gas prepared in the state with the scalar Wigner function of the one-particle density matrix of the form \eqref{Wign_func_scalar_MB} is paramagnetic \cite{Delsante1979,Anand1981}. Furthermore, there is an anti-screening of the magnetic field produced by a stationary current with the typical anti-screening radius
\begin{equation}
    r_{asc}:=(2\sqrt{1-\vk}\s)^{-1}.
\end{equation}
For $r\lesssim r_{asc}/5$, we have
\begin{equation}
    G(\spx) \approx \frac{1}{4\pi r} \big(1 +\frac{\vk}{1-\vk}\frac{r}{r_{asc}} \big),
\end{equation}
i.e., the effective current creating the magnetic field in the neutron gas grows linearly with $r$ for small $r$. For example, for $\s$ corresponding to the temperature $1$ K and $\vk\ll1$, we have $r_{asc}\approx 2.2$ nm. For $\vk\rightarrow1-0$, the anti-screening radius tends to infinity that confirms once again the presence of instability in a neutron gas when condition \eqref{instab_cond_stat_1} is fulfilled. In this limit,
\begin{equation}
    G(\spx) \approx \frac{1}{4\pi r} -\frac{\s^2 r}{2\pi}+const.
\end{equation}
This Green function resembles the Cornell potential of a quark-antiquark interaction \cite{Eichten1978}. Note, however, that as we shall show below the parameter $\vk\ll1$ even for a nonequilibrium rarefied neutron gas and the aforementioned instability is not realized.

Now we turn to the properties of the solutions to the electrostatic equation \eqref{Max_eqs_stat_00} that becomes
\begin{equation}\label{Max_eq_stat_long}
    \spk^2\big[1+\chi_\parallel(|\spk|)\big]A^0(\spk)=j^0(\spk).
\end{equation}
The static longitudinal electric susceptibility \eqref{chi_par_stat} is positive and bounded (see Fig. \ref{Phi_Chi_stat_MB_plots}). Hence, Eq. \eqref{Max_eq_stat_long} always possesses a solution. The Green function for Eq. \eqref{Max_eq_stat_long} has the form
\begin{equation}\label{Green_func_long_0}
    G(\spx)=\int\frac{d\spk}{(2\pi)^3} \frac{e^{i\spk\spx}}{\spk^2[1+\chi_\parallel(|\spk|)]}=\frac{1}{4\pi\e_\parallel(0) r} +\frac{1}{4\pi^2ir} \int_{-\infty+i0}^{\infty+i0} \frac{dx}{x}\frac{e^{2\sqrt{2}ix\s r}}{1+\chi_\parallel(x)},
\end{equation}
where $x=|\spk|/(2\sqrt{2}\s)$. We shall obtain the approximate expression for this Green function by joining the asymptotics of the electric susceptibility  \eqref{chi_stat_large_arg}, \eqref{chi_stat_small_arg} for large and small momenta. Then we have approximately (see Fig. \ref{Phi_Chi_stat_MB_plots})
\begin{equation}\label{chi_k_approx}
    \chi_\parallel(x)\approx \chi_\parallel(0) \frac{1+2x^2/3}{1+4x^4/9},
\end{equation}
and the integral in \eqref{Green_func_long_0} is readily performed by residues. Inasmuch as $\chi_\parallel(0)\ll1$, we obtain approximately
\begin{equation}\label{Green_func_long0}
    G(\spx)\approx \frac{1}{4\pi\e_\parallel(0) r}\Big[1 + \sqrt{2}\chi_\parallel(0) \cos\big(\sqrt{6}\s r+\pi/4\big) e^{-\sqrt{6}\s r}\Big].
\end{equation}
In other words, there exists a small screening of a charge placed in a dilute neutron gas. Furthermore, there are small oscillations of the electric field strength that drop exponentially to zero at large distances from a charge. For example, for $\s$ corresponding to the effective temperature $1$ K, the amplitude of oscillations of the effective charge decreases by a factor of $e$ at the distance of order $0.28$ nm.

\subsection{Longitudinal plasmon-polaritons}\label{Long_Plams_Polar}

Let us now investigate the properties of longitudinal solutions to the free effective Maxwell equations \eqref{Max_eq_eff} in more detail. The dispersion law of these modes is determined by Eq. \eqref{longitud_disp_law}. This equation is not solvable analytically for arbitrary values of parameters entering into it. Therefore, we shall consider some particular cases and obtain the approximate expressions for the dispersion law of longitudinal plasmon-polaritons in a rarefied neutron gas. Furthermore, we shall analyze numerically the solutions to Eq. \eqref{longitud_disp_law} to substantiate the analytical results.

\begin{figure}[tp]
\centering
\includegraphics*[width=0.45\linewidth]{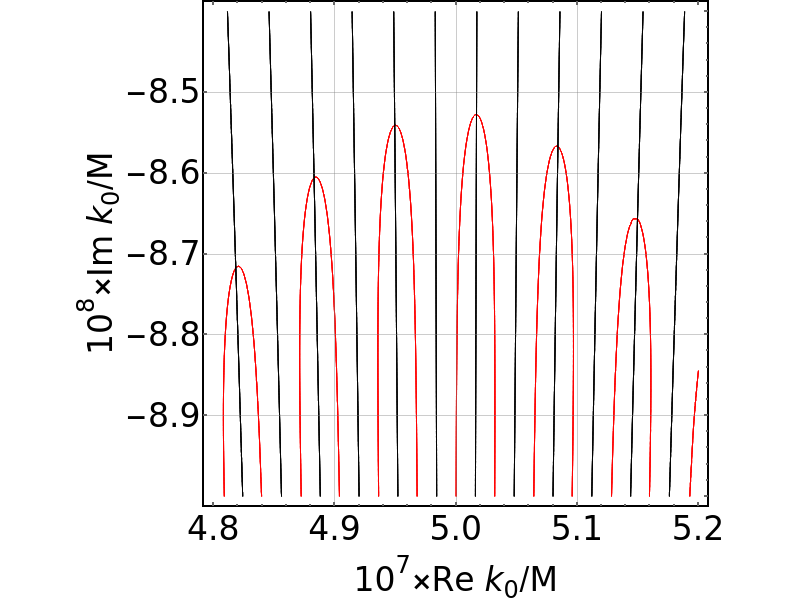}\;
\includegraphics*[width=0.45\linewidth]{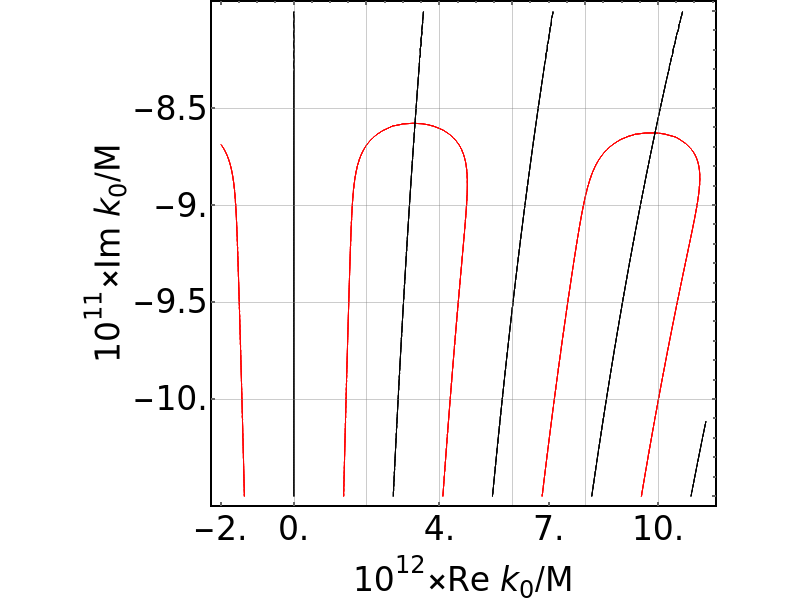}
\caption{{\footnotesize The numerical solution of Eq. \eqref{longitud_disp_law} determining the dispersion law of longitudinal plasmon-polaritons for a neutron gas with Gaussian one-particle density matrix in the complex $k_0$ plane. The energy is measured in the units of the neutron rest energy $M=0.940$ GeV. The red lines are the lines of zero real part of Eq. \eqref{longitud_disp_law}. The black lines are the lines of zero imaginary part of Eq. \eqref{longitud_disp_law}. The roots of Eq. \eqref{longitud_disp_law} are the intersection points of these lines. Left panel: The particle number density $\rho(x)=10^{-15}\rho_0= 1.2\times 10^{23}$ cm$^{-3}$, the effective temperature $\be^{-1}=10^3$ K corresponding to the rms of momenta $\s=9.57\times10^{-6} M=9.00$ keV, the respective dimensionless chemical potential $\tmu=-4.61$, the minimum energy and momentum following from \eqref{collision_conds} are $k_0^{min}= 5.18\times 10^{-19}M=4.87\times 10^{-10}$ eV and $|\spk^{min}|=5.42\times 10^{-14}M=5.10\times 10^{-5}$ eV, respectively. The plasmon-polariton momentum is $|\spk|=10^{-3}M=940$ keV. The roots of Eq. \eqref{longitud_disp_law} are well approximated by \eqref{longitud_disp_law_case1_0}, \eqref{y_longitud}. The real part of the energy of plasmon-polaritons closest to the real axis is much smaller than its imaginary part. Right panel: The same as on the left panel but for the plasmon-polariton momentum $|\spk|=10^{-6}M=940$ eV. The roots of Eq. \eqref{longitud_disp_law} are well approximated by \eqref{k_0_longit}, where the odd branches of the Lambert function should be taken.}}
\label{NumSol_Long_MB_plots}
\end{figure}

We begin with the case where
\begin{equation}\label{case1_long_plasm_pol}
   |\spk|\gg\s,\qquad|x_+|\gtrsim 1,\qquad |x_+ e^{-x_+^2}|\gg 1,
\end{equation}
and, for brevity, the notation has been introduced
\begin{equation}\label{x_pm_defn}
    x_\pm:=\frac{\al_\pm}{\sqrt{2}|\spk|\s}.
\end{equation}
In this case, the dispersion law of plasmon-polaritons is close to the dispersion law of plasmons \eqref{plasmon_disp_law}. Hence,
\begin{equation}
    x_-\approx \frac{|\spk|}{\sqrt{2}\s}\gg1,
\end{equation}
and $|x_+|\ll|x_-|$. Consequently, assuming that $\im k_0<0$, we can employ the asymptotic expansion \eqref{F_asympt_3} in the expression for
\begin{equation}
    \Phi(k)=\frac{\mu^2_p(x)
    M}{\sqrt{2}|\spk|\s} \Big[ \frac{F(x_+)}{x_+}  -\frac{F(x_-)}{x_-} \Big].
\end{equation}
So long as $|x_+|\ll|x_-|$, we can neglect the second term in the square brackets and, in virtue of the last condition in \eqref{case1_long_plasm_pol}, only the term with exponent can be retained in the expansion \eqref{F_asympt_3} for $F(x_+)$. Then
\begin{equation}\label{Phi_case1}
    \Phi(k)\approx \frac{\mu^2_p(x)M}{|\spk|\s}\sqrt{2\pi}i e^{-x_+^2}.
\end{equation}
Since $\mu_p^2(x)/M\ll1$, the dispersion law of longitudinal plasmon-polaritons \eqref{longitud_disp_law} takes the form
\begin{equation}\label{longitud_disp_law_case1}
    e^{M^2y^2/(2 \spk^2\s^2)}=\frac{ 4iM\s}{\sqrt{2\pi}\mu_p^2(x)|\spk|},
\end{equation}
where the notation has been introduced
\begin{equation}\label{longitud_disp_law_case1_0}
    -iy:=k_0-\spk^2/(2M).
\end{equation}
As a result, we come to
\begin{equation}\label{y_longitud}
    y= \frac{|\spk|\s}{M}\sqrt{2\ln \frac{4iM\s}{\sqrt{2\pi}\mu_p^2(x)|\spk|}}.
\end{equation}
As is seen, the conditions \eqref{case1_long_plasm_pol} are satisfied provided that
\begin{equation}
    \frac{4M\s}{\sqrt{2\pi}\mu_p^2(x)|\spk|}\gtrsim1.
\end{equation}
It also follows from expression \eqref{y_longitud} that there is an infinite set of longitudinal plasmon-polaritons with the same momentum $\spk$ corresponding to a different choice of the branches of the logarithm in \eqref{y_longitud} (see Fig. \ref{NumSol_Long_MB_plots}). Slowly damped modes appear for such values of \eqref{y_longitud} where the imaginary part of the energy, $-\re y$, is minimal. It is not difficult to see that the imaginary part of the energy of a longitudinal plasmon-polariton mode is much less than its real part for $|\spk|\gtrsim 100\s$. Therefore, these perturbations of the electromagnetic field can be interpreted as unstable quasiparticles. In increasing $|\spk|$, the magnitude of the ratio of the imaginary part of the energy to its real part decreases since the real part grows as $\spk^2$ whereas the imaginary part is proportional to $|\spk|$ with slow logarithmic scaling.


Notice that if the Fermi-Dirac distribution \eqref{Wign_func_scalar_FD} is considered instead of the Gaussian one-particle density matrix \eqref{Wign_func_scalar_MB} for the particle number density \eqref{dens_neutr_gas} corresponding to a nondegenerate Fermi gas, $\tmu\ll-1$, then the values of energy found above for the longitudinal plasmon-polaritons lie out of the range of applicability of the Maxwell-Boltzmann distribution (see for details Appendix \ref{Plasm-Polar_NonDeg_Fermi_App}). Nevertheless, the plasmon-polaritons with the dispersion law close to $k_0=\spk^2/(2M)$ exist even in an equilibrium nondegenerate Fermi gas but with other expression for the correction \eqref{y_longitud}. These corrections agree by the order of magnitude that corroborates the robustness of our results.

In the case when the imaginary part of the energy of a longitudinal plasmon-polariton mode, $-\re y$, is close to zero, viz., for
\begin{equation}\label{long_pp_wo_im_cond}
    \frac{4M\s}{\sqrt{2\pi}\mu_p^2(x)|\spk|}\lesssim1,\qquad |\spk|\gg\s,
\end{equation}
the term $-1$ dominates in the asymptotic expansion \eqref{F_asympt_3} for $F(x_+)$. Then Eq. \eqref{longitud_disp_law} determining the dispersion law does not depend on $\s$ and, consequently, the regime studied in \cite{ComptNeutr} is reproduced. The dispersion law of longitudinal plasmon-polariton modes is written as
\begin{equation}\label{disp_law_longit}
    k_0=\sqrt{M^2-M\mu_p^2(x)+\spk^2}-\sqrt{M^2-M\mu_p^2(x)}.
\end{equation}
For a fixed momentum $\spk$, aside from this solution to Eq. \eqref{longitud_disp_law}, there is an infinite set of solutions of the form \eqref{longitud_disp_law_case1_0} with $y$ given by formula \eqref{y_longitud}, where only those values of $y$ should be kept that have $\re y>0$ (see Fig. \ref{NumSol_Long_MB_plots}). The imaginary correction to the solution \eqref{disp_law_longit} is exponentially suppressed. Notice that condition \eqref{long_pp_wo_im_cond} cannot be fulfilled for a dilute neutron gas even in the nonequilibrium regime. Indeed, if one substitutes the value of $\mu_p^2$ into \eqref{long_pp_wo_im_cond}, takes into account that the density of a rarefied neutron gas is less than the nuclear one,
\begin{equation}
    \rho_0=\frac{3}{4\pi r_0^3},\qquad r_0=6/M,
\end{equation}
and keeps in mind that for a gas of fermions $\s^2\gtrsim M/\beta_F$, where $\beta_F$ is the reciprocal degeneracy temperature \eqref{nondeg_gas_cond}, then condition \eqref{long_pp_wo_im_cond} holds only for the photon momenta $|\spk|\gg3.6\times 10^3 M$, i.e., out of the range of applicability of the approximations we use. It is not difficult to verify that condition \eqref{long_pp_wo_im_cond} cannot be satisfied for a Gaussian wave packet of a single neutron, either.

Now we turn to the case $|\spk|\ll\s$. Then
\begin{equation}
    x_--x_+=-\frac{k_0^2-\spk^2}{\sqrt{2}|\spk|\s}\approx \frac{|\spk|}{\sqrt{2}\s}\ll1,
\end{equation}
where it has been taken into account in the approximate equality that $|k_0|\ll|\spk|$. Therefore,
\begin{equation}\label{Phi_k_k_ll_s}
    \Phi(k)\approx -\frac{\vk}{2} \Big(\frac{F(x)}{x}\Big)',
\end{equation}
where $x=(x_+ +x_-)/2=Mk_0/(\sqrt{2}|\spk|\s)$. Inasmuch as
\begin{equation}
    \Big(\frac{F(x)}{x}\Big)'=-2(1+F(x)),
\end{equation}
the asymptotic expansion \eqref{F_asympt_3} can be used for $|x|\gg1$ and $\im k_0<0$, where the term with exponent dominates. Then the equation for the dispersion law of longitudinal plasmon-polariton modes \eqref{longitud_disp_law} is approximately reduced to
\begin{equation}
     ixe^{(ix)^2} =-\frac{M}{4\sqrt{\pi}\mu_p^2(x)}.
\end{equation}
As a result, we arrive at the infinite set of solutions
\begin{equation}\label{k_0_longit}
    k_0 =-i\frac{|\spk|\s}{M} \sqrt{ W\Big(\frac{1}{8\pi\chi^2_\parallel(0)}\Big)},
\end{equation}
where $W(z)$ is the Lambert function. The different branches of the Lambert function can be taken in expression \eqref{k_0_longit}. This gives rise to an infinite set of solutions of the dispersion equation \eqref{longitud_disp_law} in this regime (see Fig. \ref{NumSol_Long_MB_plots}), where only such solutions should be picked out that $\im k_0<0$. These solutions describe rapidly decaying resonances with the linear dispersion law in the domain of momenta $|\spk|\ll\s$.

\subsection{Transverse plasmon-polaritons}\label{Trans_Plasm_Polar}

\begin{figure}[tp]
\centering
\includegraphics*[width=0.45\linewidth]{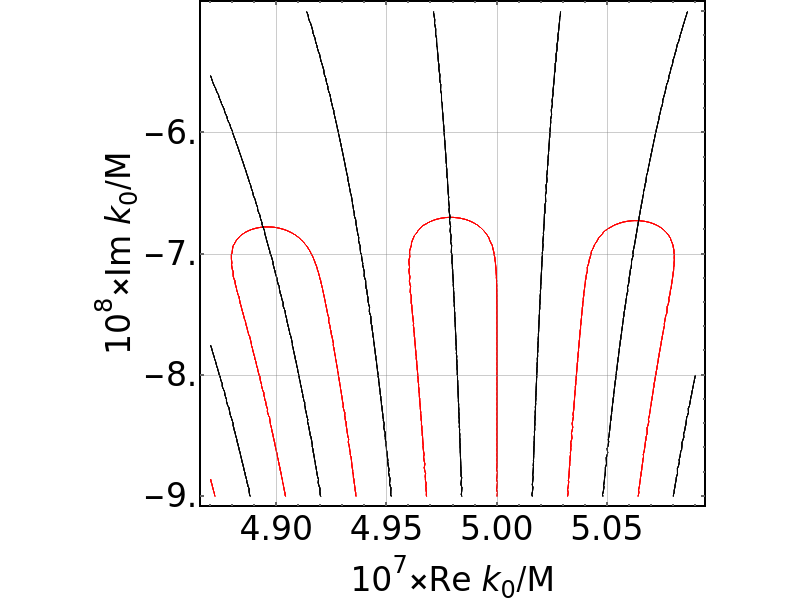}\;
\includegraphics*[width=0.45\linewidth]{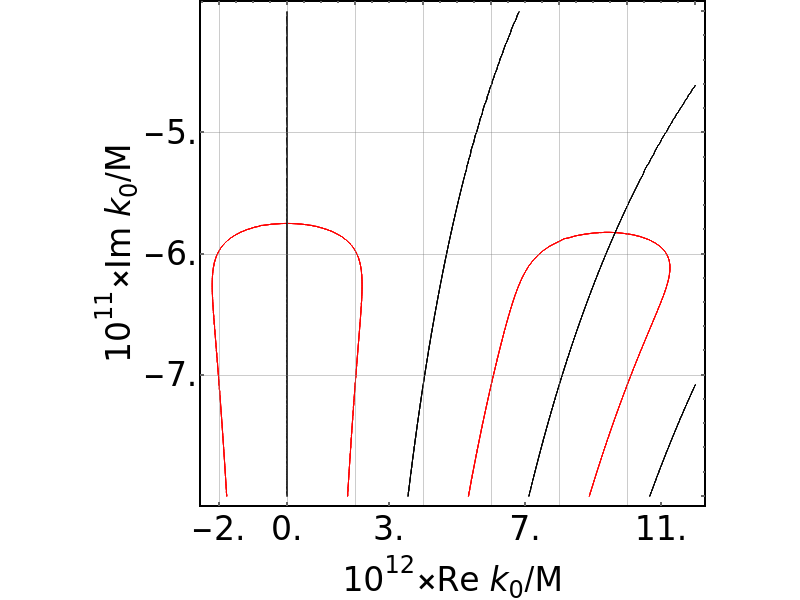}
\caption{{\footnotesize The same as in Fig. \ref{NumSol_Long_MB_plots} but for Eq. \eqref{transv_disp_law} determining the dispersion law of transverse plasmon-polaritons for a neutron gas with Gaussian one-particle density matrix in the complex $k_0$ plane. Left panel: The parameters are the same as on the left panel in Fig. \ref{NumSol_Long_MB_plots}. The roots of Eq. \eqref{transv_disp_law} are well approximated by \eqref{longitud_disp_law_case1_0}, \eqref{y_transv}. The real part of the energy of plasmon-polaritons closest to the real axis is much smaller than its imaginary part. Right panel: The same as on the left panel but for the plasmon-polariton momentum $|\spk|=10^{-6}M=940$ eV. The roots of Eq. \eqref{transv_disp_law} are well approximated by \eqref{k_0_trans_k_less_sigma}, where the even branches of the Lambert function should be taken.} }
\label{NumSol_Trans_MB_plots}
\end{figure}

Let us carry out the analogous analysis of transverse plasmon-polariton modes with the dispersion law determined by Eq. \eqref{transv_disp_law}. As in the previous section, we consider, at first, the region of parameters \eqref{case1_long_plasm_pol} where the function $\Phi(k)$ takes approximately the form \eqref{Phi_case1}. Introducing the notation \eqref{longitud_disp_law_case1_0}, we come to the equation
\begin{equation}\label{transv_disp_law_case1}
    e^{M^2y^2/(2 \spk^2\s^2)}=\frac{ |\spk|\s}{\sqrt{2\pi}i\mu_p^2(x) M}.
\end{equation}
Hence,
\begin{equation}\label{y_transv}
    y =\frac{|\spk|\s}{M}\sqrt{2\ln \frac{ |\spk|\s}{\sqrt{2\pi}i\mu_p^2(x) M}}.
\end{equation}
This gives the infinite set of values of energy $k_0$ of plasmon-polaritons at a fixed momentum $\spk$. The conditions \eqref{case1_long_plasm_pol} are satisfied provided that
\begin{equation}
    \frac{ |\spk|\s}{\sqrt{2\pi}\mu_p^2(x) M}\gtrsim1.
\end{equation}
The numerical solution of Eq. \eqref{transv_disp_law} in this domain of parameters is presented in Fig. \ref{NumSol_Trans_MB_plots}. As in the case of longitudinal modes, these plasmon-polaritons can be interpreted as unstable quasiparticles for $|\spk|\gtrsim 100\s$. Furthermore, just as in the case of longitudinal modes, expression \eqref{y_transv} is not valid for a nondegenerate Fermi gas obeying the Fermi-Dirac distribution with the dimensionless chemical potential, $\tmu\ll-1$, determined by Eq. \eqref{dens_neutr_gas}. Nevertheless, even in the case of an equilibrium nondegenerate Fermi gas, there are the transverse plasmon-polaritons with the dispersion law close to $k_0=\spk^2/(2M)$ (see Appendix \ref{Plasm-Polar_NonDeg_Fermi_App}). The correction to the dispersion law $k_0=\spk^2/(2M)$ coincides by the order of magnitude with \eqref{y_transv} confirming the robustness of the results.


For
\begin{equation}\label{transv_disp_law_wo_im}
    \frac{ |\spk|\s}{\sqrt{2\pi}\mu_p^2(x) M}\lesssim1,\qquad |\spk|\gg\s,
\end{equation}
just as in the previous section, we have that in this parameter domain the dispersion law of transverse plasmon-polariton modes is well described by the expressions derived in \cite{ComptNeutr}. Formula (118) of \cite{ComptNeutr} implies approximately
\begin{equation}\label{imaginary_root}
    k^2_0\approx \frac{\spk^2}{2M}\Big[\frac{\spk^2}{2M} -2\mu^2_p(x)\Big].
\end{equation}
It is clear that for
\begin{equation}
    \s\ll|\spk|<2\sqrt{M\mu^2_p(x)},
\end{equation}
the transverse plasmon-polariton modes are unstable and grow exponentially with time \cite{ComptNeutr}.

In the region of photon momenta $|\spk|\ll\s$, the function $\Phi(k)$ has the form \eqref{Phi_k_k_ll_s}. On substituting this expression into Eq. \eqref{transv_disp_law} determining the dispersion law of transverse plasmon-polaritons, we formally obtain
\begin{equation}\label{transv_form_sol}
    k_0=-i\frac{\sqrt{2}|\spk|\s}{M} F^{-1}\big(\vk^{-1}-1\big),
\end{equation}
where $F^{-1}(x)$ is the multivalued function inverse to \eqref{plasm_funk}. In the case $|x|\gg1$, equation \eqref{transv_disp_law} is reduced to
\begin{equation}
     ix e^{(ix)^2}=\frac{1}{2\sqrt{\pi}\vk},
\end{equation}
whence we have
\begin{equation}\label{k_0_trans_k_less_sigma}
    k_0=-i\frac{|\spk|\s}{M} \sqrt{W\Big(\frac{1}{2\pi\vk^2} \Big)},
\end{equation}
i.e., we obtain an infinite set of rapidly decaying resonances with a linear dispersion law. For $|x|\ll 1$, using expansion  \eqref{F_small_arg}, expression \eqref{transv_form_sol} gives approximately
\begin{equation}\label{k_0_trans_k_less_sigma_small_x}
    k_0=-i\sqrt{\frac{2}{\pi}}\frac{|\spk|\s}{M} \frac{\vk^{-1}-1}{1+2(\vk^{-1}-1)/\pi}.
\end{equation}
It is obvious that there is an instability when condition \eqref{instab_cond_stat_1} holds. The transverse modes with momenta $|\spk|<2\s\vk^{1/2}$ are unstable for $\vk\geqslant1$, which can be interpreted as the existence of a spontaneous magnetization of a neutron gas due to anomalous magnetic moments of neutrons. In other words, the neutron gas in this region of parameters becomes ferromagnetic.


\begin{figure}[tp]
\centering
\includegraphics*[width=0.45\linewidth]{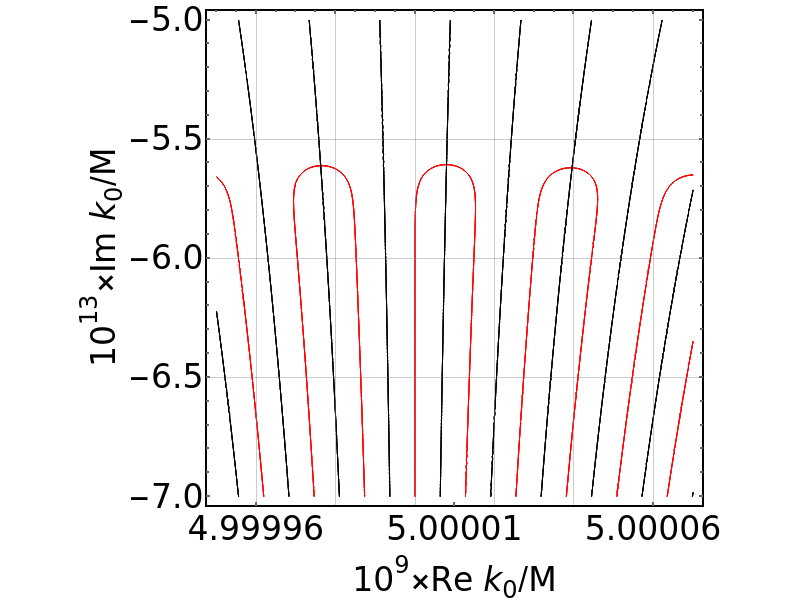}\;
\includegraphics*[width=0.45\linewidth]{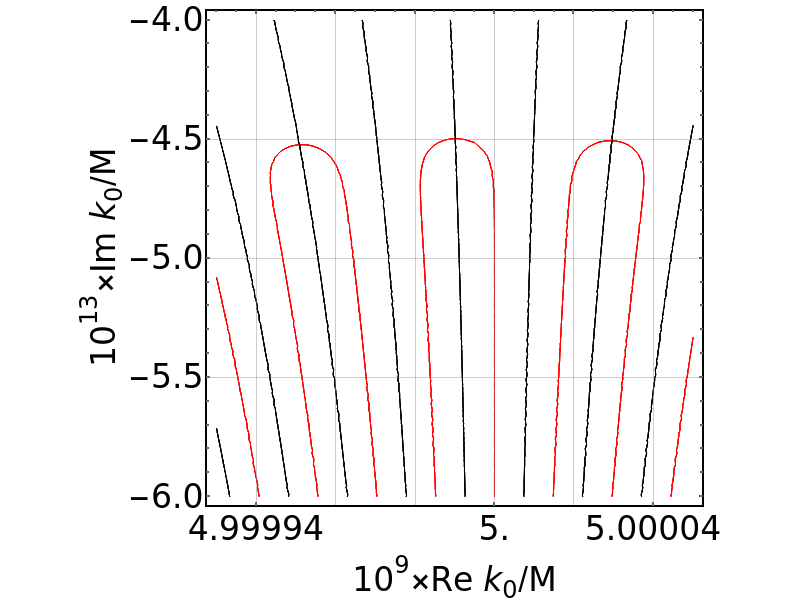}
\caption{{\footnotesize The same as in Figs. \ref{NumSol_Long_MB_plots}, \ref{NumSol_Trans_MB_plots} but for the Gaussian wave packet \eqref{Wign_func_scalar_MB} of a single neutron with $\s=5.32\times 10^{-10} M=0.5$ eV and $\s_x=1/(2\s)=0.20$ $\mu$m. The plasmon-polariton momentum is $|\spk|=10^{-4}M=94$ keV. It is seen that the relative magnitude of the imaginary part of the energy of plasmon-polaritons is much less than its real part and so these plasmon-polaritons can be interpreted as unstable quasiparticles. Left panel: The solution of Eq. \eqref{longitud_disp_law} for longitudinal plasmon-polaritons. The roots of Eq. \eqref{longitud_disp_law} are well approximated by \eqref{longitud_disp_law_case1_0}, \eqref{y_longitud}. Right panel: The solution of Eq. \eqref{transv_disp_law} for transverse plasmon-polaritons. The roots of Eq. \eqref{transv_disp_law} are well approximated by \eqref{longitud_disp_law_case1_0}, \eqref{y_transv}.} }
\label{NumSol_MB_1neutr_plots}
\end{figure}


However, inequality \eqref{instab_cond_stat_1} cannot be fulfilled for a single neutron \cite{ComptNeutr}. It cannot be fulfilled for a rarefied neutron gas even in a nonequilibrium state, either. Indeed, we have for fermions
\begin{equation}
    \s^2\gtrsim\frac{M}{\be_F}=\frac12 (3\pi^2\rho(x))^{2/3}.
\end{equation}
For condition \eqref{instab_cond_stat_1} to be satisfied, it is necessary
\begin{equation}\label{instab_cond_stat_MB_1}
    \mu_p^2(x)M \gtrsim \frac12 (3\pi^2\rho(x))^{2/3},
\end{equation}
which is equivalent to
\begin{equation}\label{rho_instab}
    \rho(x)\gtrsim \frac{9\pi^4}{8(M\mu_p^2)^3}\approx \frac{81 \pi^2}{16\al^3}\rho_0,
\end{equation}
where we have taken into account that $\mu_p^2\approx 4\pi\al/M^2$. The inequality \eqref{rho_instab} does not hold for a low-density neutron gas. Conducting reasoning along the same lines, it can be shown that the domain of parameters \eqref{transv_disp_law_wo_im} is not reachable for a neutron gas and for a single neutron.

In our considerations we have not taken into account the neutron collisions. Let us estimate the domain of applicability of such an approximation. The collision frequency equals
\begin{equation}
    \nu=\rho(x)\s_{nn}\ups,\qquad \ups\sim\s/M,
\end{equation}
where the total cross section of neutron-by-neutron scattering, $\s_{nn}$, has been introduced. For low energies that we consider, it is of order (see, e.g., \cite{Mitchell2005,Chatziioannou2025,Vidana2021,Gandolfi2015,Gezerlis2010})
\begin{equation}\label{sigma_nn}
    \s_{nn}\sim 2.2\times 10^3\,\text{fm}^2\approx 4.9\times10^4 M^{-2}.
\end{equation}
Notice that the electromagnetic contribution to the total cross section of neutron-neutron scattering due to the presence of an anomalous magnetic moment is of order $\s^{em}_{nn}\sim\mu_p^4 M^2\approx 3.7\times 10^{-4}$ fm${}^2$. It is negligibly small in comparison with the nuclear contribution \eqref{sigma_nn}. The solutions to the effective Maxwell equations obtained in this and the previous sections make sense only if
\begin{equation}\label{collision_conds}
    |k_0|\gg\nu\sim \rho(x)\s_{nn}\s/M,\qquad |\spk|\gg \nu/(\s/M)\sim\rho(x)\s_{nn}.
\end{equation}
Under the fulfillment of these conditions, one can neglect collisions in a neutron gas or take them into account as a small correction. It is clear that, in describing the plasmon-polariton modes on the wave packet of a single neutron, conditions \eqref{collision_conds} do not need to be imposed. The numerical solutions of Eqs. \eqref{longitud_disp_law}, \eqref{transv_disp_law} determining the dispersion laws of longitudinal and transverse plasmon-polaritons on the Gaussian wave packet of a single neutron are presented in Fig. \ref{NumSol_MB_1neutr_plots}.

\section{Neutron gas in a thermodynamic equilibrium}\label{Neutrons_FD_Distr}

Let us consider the case where the scalar part of the Wigner function of the one-particle density matrix has the form \eqref{Wign_func_scalar_FD}, i.e., a rarefied neutron is in a thermodynamic equilibrium. We suppose that the nonrelativistic approximation is valid, viz., conditions \eqref{k_nonrel}, \eqref{nonrel_appr_2} are fulfilled, where one ought to substitute $\s$ in the form \eqref{sigma_subs_FD}.

The general expression for the transverse part of the polarization operator is given in \eqref{Pi_par_Pi_perp_gen}. In the nonrelativistic approximation, it can be cast into the form
\begin{equation}
    \Pi_\perp \approx \mu_p^2M k^2 \int d\spp \rho(x,\spp) \Big(\frac{1}{\al_- -\spk\spp} -\frac{1}{\al_+ -\spk\spp}\Big).
\end{equation}
Bearing in mind that
\begin{equation}\label{Li0}
    \frac{1}{e^{\be(\spp^2/(2M)-\mu)}+1}=-\Li_{0}(-e^{\be\mu-\be\spp^2/(2M)}),
\end{equation}
the integral over the momentum can be performed. Decomposing the momentum $\spp$ into the longitudinal and transverse parts to the vector $\spk$ and integrating over the transverse components with the aid of formula \eqref{polylog_mellin_transf}, we deduce
\begin{equation}\label{Pi_perp_FD}
    \Pi_\perp = \mu^2_p(x)
    M k^2 \Big[\frac{1}{\alpha_+} F^{(1)}_D\Big(\be\mu,\frac{\sqrt{\be}\alpha_+}{\sqrt{2M}|\spk|}\Big) -\frac{1}{\alpha_-} F^{(1)}_D\Big(\be\mu,\frac{\sqrt{\be}\alpha_-}{\sqrt{2M}|\spk|}\Big) \Big],
\end{equation}
where $\mu^2_p(x):=\mu_p^2\rho(x)$ and $\rho(x)$ is the particle number density of neutrons \eqref{dens_neutr_gas}. The definition and some properties of the function $F^{(\nu)}_D(\tmu,x)$ are presented in Appendix \ref{App_F_D_Expansions}. In what follows, for uniformity of notation, we denote $\s\equiv\sqrt{M/\beta}$ and
\begin{equation}\label{Phi_k_FD}
    \Phi(k):=\mu^2_p(x)
    M\Big[ \frac{1}{\alpha_+} F_D^{(1)}\Big(\tmu,\frac{\alpha_+}{\sqrt{2}|\spk|\s}\Big) -\frac{1}{\alpha_-}F_D^{(1)}\Big(\tmu,\frac{\alpha_-}{\sqrt{2}|\spk|\s}\Big) \Big],
\end{equation}
where $\tmu=\be\mu$. Then the magnetic permeability of transverse modes has the form \eqref{mu_perp_Phi}.

The expression for the longitudinal part of the photon polarization operator was also given in formula \eqref{Pi_par_Pi_perp_gen}. In the nonrelativistic approximation, using relations \eqref{Li0} and \eqref{polylog_mellin_transf}, the integral over the momenta can be evaluated. As a result,
\begin{equation}\label{chi_parallel_FD}
\begin{split}
    \chi_\parallel(k)=-\frac{\Pi_\parallel}{k^2} =\,&-\frac{\mu^2_p(x)}{M}
    \bigg\{1+\frac{k^2}{4} \Big[\frac{1}{\alpha_+} F_D^{(1)}\Big(\tmu,\frac{\alpha_+}{\sqrt{2}|\spk|\s}\Big) -\frac{1}{\alpha_-}F_D^{(1)}\Big(\tmu,\frac{\alpha_-}{\sqrt{2}|\spk|\s}\Big)\Big]-\\ &-2\s^2\frac{\Li_{5/2}(-e^{\tmu})}{\Li_{3/2}(-e^{\tmu})} \Big[\frac{1}{\alpha_+} F_D^{(2)}\Big(\tmu,\frac{\alpha_+}{\sqrt{2}|\spk|\s}\Big) -\frac{1}{\alpha_-}F_D^{(2)}\Big(\tmu,\frac{\alpha_-}{\sqrt{2}|\spk|\s}\Big)\Big] \bigg\}.
\end{split}
\end{equation}
The longitudinal dielectric permittivity equals $\e_\parallel=1+\chi_\parallel$.

According to the general analysis carried out in Sec. \ref{Pol_Oper_LDNG}, the Maxwell equations \eqref{Max_eq_eff} have the solutions in the form of longitudinal and transverse plasmon-polariton modes. The dispersion law of longitudinal plasmon-polaritons is determined by Eq. \eqref{longitud_disp_law}. The transverse plasmon-polaritons have the dispersion law following from Eq. \eqref{transv_disp_law}. As long as $F^{(\nu)}_D(\tmu,x)$ obeys the symmetry relation \eqref{F_D_symm}, the dispersion law of plasmon-polariton modes, $k_0(\spk)$, possesses the symmetry discussed in Sec. \ref{Pol_Oper_Gauss} after formula \eqref{longitud_disp_law}.

As is discussed in detail in Appendix \ref{App_F_D_Expansions}, the Fermi-Dirac distribution goes into the Maxwell-Boltzmann one for $\tmu\ll-1$, while the transverse \eqref{Pi_perp_FD} and longitudinal \eqref{chi_parallel_FD} parts of the polarization operator are reduced to \eqref{Pi_perp_Max} and \eqref{Pi_parallel_Max}, respectively. Therefore, we shall mainly be interested in the electromagnetic properties of a low-density neutron gas in the state close to a degenerate Fermi gas. In the degenerate state, we have \cite{LandLifStatPhysP1}
\begin{equation}
    \rho(x)=\frac{p_F^3}{3\pi^2},\qquad\mu=\be_F^{-1}=\frac{p_F^2}{2M},
\end{equation}
and the functions $F_D^{(1,2)}(\tmu,x)$ turn into \eqref{F_D_degen0}, where
\begin{equation}
    z_\pm=\frac{\al_\pm}{\sqrt{2\tmu}|\spk|\s}=\frac{\al_\pm}{p_F|\spk|},
\end{equation}
and
\begin{equation}
    2\s^2\frac{\Li_{5/2}(-e^{\tmu})}{\Li_{3/2}(-e^{\tmu})}\approx\frac{2}{5}p_F^2.
\end{equation}
Of course, expressions \eqref{Pi_perp_FD} and \eqref{chi_parallel_FD} with the functions $F_D^{(1,2)}(\tmu,x)$ in the form \eqref{F_D_degen0} can be obtained from the general formula \eqref{Pi_par_Pi_perp_gen}, where the Wigner function of the one-particle density matrix is taken in the form of the Fermi-Dirac step.

\subsection{Static limit}\label{Stat_Lim_FD}

We start the investigation of solutions to the effective Maxwell equations \eqref{Max_eq_eff} with the static limit for the transverse modes, viz., with the solutions to Eq. \eqref{Max_eq_stat_transv0}. In this limit, the approximate equality \eqref{alpha_pm_stat} is valid and we have (see also \cite{Bobrov2018})
\begin{equation}\label{Phi_FD_k_stat}
    \Phi(\spk)\approx -\frac{\vk}{4} \frac{F^{(1)}_D(\tmu,x+i0)+F^{(1)}_D(\tmu,-x+i0)}{x^2}\Big|_{x=|\spk|/(2\sqrt{2}\s)},
\end{equation}
where $\vk=\be\mu_p^2(x)$. It is not difficult to prove that $\Phi(\spk)>0$ for $x\in \mathbb{R}$. Moreover, the numerical analysis reveals that $\Phi(k)$ possesses a maximum at $k=0$ and decreases monotonically for $k>0$. It follows from \eqref{Phi_FD_k_stat} and \eqref{F_D_expans_zero} that
\begin{equation}
    \Phi(0)=\vk \frac{\Li_{1/2}(-e^{\tmu})}{\Li_{3/2}(-e^{\tmu})}.
\end{equation}
In the case of a degenerate neutron gas \cite{Clark1969},
\begin{equation}\label{Phi_0_degen}
    \Phi(0)=\frac{3\mu_p^2(x)}{2\mu}=\frac{\mu_p^2 M^2}{\pi^2}\frac{p_F}{M}\ll1,
\end{equation}
where we have used the asymptotic expansion of the polylogarithm \eqref{polylog_large_mu}. For
\begin{equation}\label{large_arg_stat}
    \be\frac{\spk^2}{8 M}\gg\max(1,\tmu),
\end{equation}
the asymptotics \eqref{Phi_k_stat_large_arg} holds. It ensues from the expansion \eqref{F_D_asympt}. As for a degenerate neutron gas, condition \eqref{large_arg_stat} is equivalent to $|\spk|\gg p_F$. The effective Maxwell equations with the stationary current density $j^i(\spk)$ have the form \eqref{Max_eq_stat_transv} with $\Phi(\spk)$ given in \eqref{Phi_FD_k_stat}. The estimates \eqref{instab_cond_stat_MB_1}, \eqref{rho_instab}, and \eqref{Phi_0_degen} imply that the instability condition, $\Phi(0)\geqslant1$, is not fulfilled for a rarefied neutron gas even in a thermodynamically nonequilibrium state.


\begin{figure}[tp]
\centering
\includegraphics*[width=0.47\linewidth]{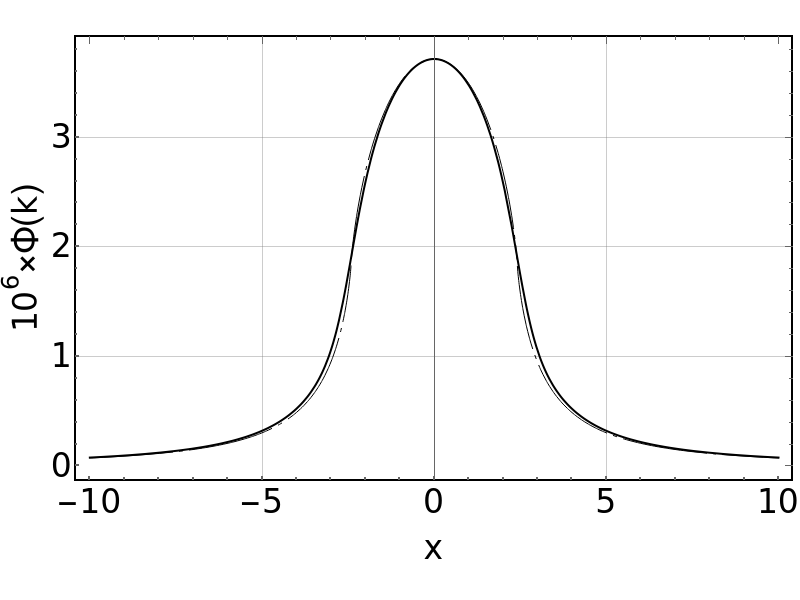}\;
\includegraphics*[width=0.48\linewidth]{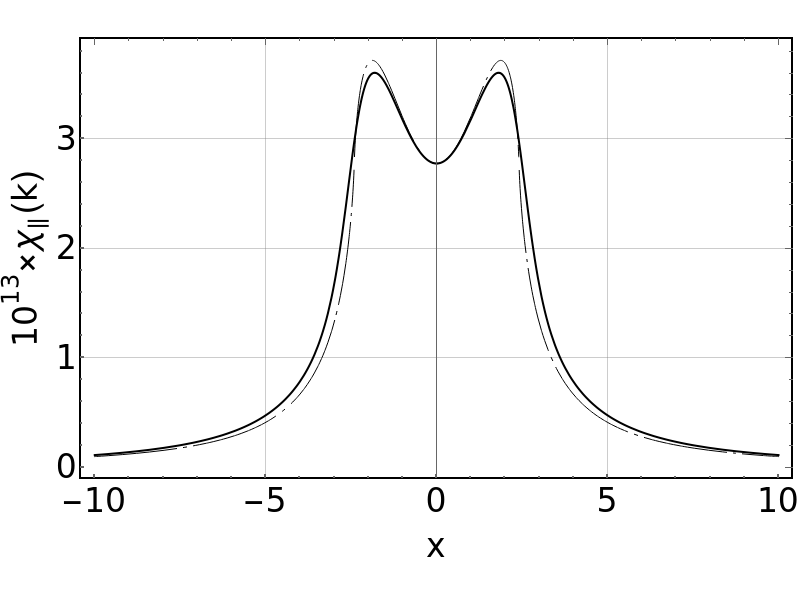}
\caption{{\footnotesize The functions $\Phi(k)$ and $\chi_\parallel(k)$ for a neutron gas obeying the Fermi-Dirac distribution \eqref{Wign_func_scalar_FD} in the static limit. Here $x=k/(2\sqrt{2}\s)$ and $x=1$ corresponds to $k=3.8\times 10^{-4}M=0.36$ MeV. The particle number density $\rho(x)=3\times10^{-9}\rho_0= 3.6\times 10^{29}$ cm$^{-3}$, the effective temperature $\be^{-1}=2\times10^5$ K, the respective dimensionless chemical potential $\tmu=5.8$, the Fermi momentum $p_F=4.6\times10^{-4}M=0.434$ MeV, the minimum momentum following from \eqref{collision_conds} is $|\spk^{min}|=1.6\times 10^{-7}M=153$ eV. The solid lines are the exact values, whereas the dashed-dotted lines are the approximations for a degenerate neutron gas \eqref{Phi_FD_k_stat_deg} and \eqref{chi_par_stat_deg}.} }
\label{Phi_chi_FD_stat_plots}
\end{figure}


Let us find the asymptotics of the Green function for Eq. \eqref{Max_eq_stat_transv} at $r\rightarrow\infty$. So long as $\Phi(k)\neq1$ on the real axis, the representation \eqref{Green_func_0} is valid. The asymptotics of the second term in \eqref{Green_func_0} for $r\rightarrow\infty$ is determined by the singular points of $1/(1-\Phi(k))$ in the upper complex half-plane. Such points can appear either at $\Phi(k)=1$ or due to the singular points \eqref{branch_points} of the function $F_D(\tmu,-x)$ entering into the definition of $\Phi(k)$. For $\vk\ll1$, the singularities of the second type \eqref{branch_points} lie closer to the real axis when $\tmu\gg1$, i.e., in the case of a degenerate Fermi gas. For the closest singular points \eqref{branch_points}, we have approximately
\begin{equation}
    k\approx \pm2p_F\big(1 \pm\frac{i\pi}{2\tmu}\big)=\pm 2p_F +\frac{2\pi iM}{\be p_F}.
\end{equation}
Then
\begin{equation}
    \frac{1}{4\pi^2ir} \int_{-\infty+i0}^{\infty+i0} \frac{dk}{k}\frac{e^{ikr}}{1-\Phi(k)}\sim \cos(2p_F r+\de)e^{-2\pi Mr/(\be p_F)},
\end{equation}
i.e., the counterpart of Friedel oscillations arises (see, e.g., Chap. 5 of \cite{Ziman1972}).

In order to find the exact form of the asymptotics of this contribution at large $r$, we consider the degenerate neutron gas, for which $F^{(1)}_D(\tmu,x)$ takes the form \eqref{F_D_degen0}, and perform the calculation of the asymptotics as it was done in Sec. 40 of \cite{LandLifPhysKin}. Notice that, in the limit of a degenerate Fermi-gas, $F^{(1)}_D(\tmu,x)$ possesses the branch points at $x=\pm\sqrt{\tmu}$ and so the $i0$-prescription given in formula \eqref{Phi_FD_k_stat} has to be used. Then for $k\in \mathbb{R}$,
\begin{equation}\label{Phi_FD_k_stat_deg}
    \Phi(k)=\frac{\Phi(0)}{2} \Big[1-\frac{1-z^2}{2z}\ln\Big|\frac{1-z}{1+z}\Big|\Big],
\end{equation}
where $z=k/(2p_F)$. The comparison of this approximation with the exact function $\Phi(k)$ is given in Fig. \ref{Phi_chi_FD_stat_plots}. It is useful to write the Green function in the case we consider as
\begin{equation}
    G(\spx)=\frac{1}{4\pi r}\frac{1}{1-\Phi(0)} +\frac{1}{4\pi^2 ir} \Big(\int_{C_-} +\int_{C_+}\Big)\frac{dk}{k}\frac{e^{ikr}}{1-\Phi(k)}.
\end{equation}
The contour $C_-$ runs from $-\infty$ slightly above the real axis, turns to the left at the point $k=-2p_F$, and goes to $+i\infty$ parallel to the imaginary axis. The contour $C_+$ is obtained from $C_-$ by a reflection in the imaginary axis and passes in the direction from $+i\infty$ to $+\infty$. Let us consider the contribution of the integral along the contour $C_-$. On the part of the contour $(-\infty,-2p_F]$, we have
\begin{equation}
    \Phi(k)=\frac{\Phi(0)}{2} \Big[1-\frac{1-z^2}{2z}\ln\frac{1-z}{-1-z}\Big];
\end{equation}
and on the part of the contour $[-2p_F,+i\infty)$, we obtain
\begin{equation}
    \Phi(k)=\frac{\Phi(0)}{2} \Big[1-\frac{1-z^2}{2z}\big(-i\pi+\ln\frac{1-z}{-1-z}\big)\Big].
\end{equation}
For $p_F r\gg1$, the main contribution to the integral along the contour $C_-$ comes from the neighborhood of the point $k=-2p_F$. Expanding $1/(k(1-\Phi(k)))$ in the vicinity of this point, we derive in the leading order
\begin{equation}
    \frac{1}{4\pi^2 ir} \int_{C_-} \frac{dk}{k}\frac{e^{ikr}}{1-\Phi(k)}\approx \frac{-i\pi}{4\pi^2 ir}\frac{e^{-2ip_Fr}}{-2p_F} \frac{a}{b^2} \int_0^{+i\infty} \frac{dkk}{-2p_F}e^{ikr}=\frac{a}{4b^2 p_F^2} \frac{e^{-2ip_Fr}}{4\pi r^3},
\end{equation}
where
\begin{equation}
    a=\Phi(0)/2,\qquad b=1-a.
\end{equation}
Performing the analogous calculations for the integral along $C_+$ and summing the resulting expressions, we arrive at
\begin{equation}\label{Green_func_trans}
    G(\spx)\approx\frac{1}{4\pi r}\frac{1}{1-\Phi(0)} +\frac{a}{2b^2p_F^2} \frac{\cos(2p_Fr)}{4\pi r^3}e^{-2\pi Mr/(\be p_F)},
\end{equation}
where $p_F r\gg1$ and we have restored the exponential factor depending on the temperature that shows the domain of applicability of the asymptotics obtained. It is seen that, in contrast to the Friedel oscillations in a degenerate electron plasma, the oscillating term in a degenerate neutron gas is subleading at large $r$.

Now we consider the solutions to the effective Maxwell equation \eqref{Max_eq_stat_long} for an electrostatic potential. As follows from \eqref{chi_parallel_FD}, the static longitudinal electric susceptibility is written as
\begin{equation}\label{chi_par_stat_FD}
    \chi_\parallel(\spk)=-\chi_\parallel(0) \Big[1+ \frac{F^{(1)}_D(\tmu,x)+F^{(1)}_D(\tmu,-x)}{2} +\frac{\Li_{5/2}(-e^{\tmu})}{\Li_{3/2}(-e^{\tmu})} \frac{F^{(2)}_D(\tmu,x)+F^{(2)}_D(\tmu,-x)}{2x^2}\Big]_{x=|\spk|/(2\sqrt{2}\s)},
\end{equation}
where $\chi_\parallel(0)=\mu_p^2(x)/M$. This expression is a generalization of formula \eqref{chi_par_stat}. If necessary, the $i0$-prescription is assumed in this expression as in \eqref{Phi_FD_k_stat}. The electric susceptibility \eqref{chi_par_stat_FD} is positive and bounded. Therefore, equation \eqref{Max_eq_stat_long} always has a solution.

Let us obtain the asymptotics of the Green function of Eq. \eqref{Max_eq_stat_long} for $r\rightarrow\infty$ in the case of a rarefied neutron gas in the state close to a degenerate Fermi gas. The evaluation of this asymptotics is carried out along the same lines as it has been done for the Green function of transverse modes \eqref{Green_func_trans}. In the limit of a degenerate Fermi gas, we have from \eqref{chi_par_stat_FD},
\begin{equation}\label{chi_par_stat_deg}
    \chi_\parallel(k)=\frac{\chi_\parallel(0)}{4} \Big[1+3z^2-\frac32\frac{1-z^4}{z}\ln\Big|\frac{1-z}{1+z}\Big| \Big],
\end{equation}
where $z=k/(2p_F)$. The comparison of the approximation \eqref{chi_par_stat_deg} with the exact function $\chi_\parallel(k)$ is presented in Fig. \ref{Phi_chi_FD_stat_plots}. Performing the calculations as above, we arrive at
\begin{equation}\label{Green_func_long_asympt}
    G(\spx)\approx\frac{1}{4\pi\e_\parallel(0) r} +\frac{3\chi_\parallel(0)}{4\e^2_\parallel(0)p_F^2} \frac{\cos(2p_Fr)}{4\pi r^3}e^{-2\pi Mr/(\be p_F)},
\end{equation}
where we have restored the exponential factor depending on the temperature. Just as in the case of the oscillations of transverse modes \eqref{Green_func_trans}, the oscillations of the electrostatic potential in a degenerate neutron gas are subleading at $r\rightarrow+\infty$.

\subsection{Plasmon-polaritons}\label{Plasm_Polar_FD}

Consider the plasmon-polariton modes in a degenerate dilute neutron gas. Let
\begin{equation}
    z_\pm=\frac{x_\pm}{\sqrt{\tmu}}=\frac{\al_\pm}{|\spk|p_F},
\end{equation}
where $x_\pm$ are defined in \eqref{x_pm_defn}. Then
\begin{equation}
    \Phi(k)=\frac{\mu_p^2Mp_F^2}{3\pi^2|\spk|}\Big[\frac{1}{z_+} F_D^{(1)}(\tmu,x_+) -\frac{1}{z_-}F_D^{(1)}(\tmu,x_-)\Big],
\end{equation}
where the expression for $F_D^{(1)}(\tmu,x)$ should be taken from \eqref{F_D_degen0}. The solutions to Eq. \eqref{transv_disp_law} determining the dispersion law of nontrivial transverse modes exist only in the case $\im z_\pm<0$ and $|z_\pm|\gg1$. In this domain,
\begin{equation}\label{F_D_1_asympt}
    F^{(1)}_D(\tmu,z_\pm)\approx -\frac{3i\pi}{2}z_\pm^3.
\end{equation}
Substituting this expression into \eqref{transv_disp_law}, we obtain
\begin{equation}
    \Big(\frac{k^2_0}{\spk^2}-1\Big)\frac{k_0}{|\spk|}=\frac{i\pi}{\mu_p^2M^2}.
\end{equation}
Consequently,
\begin{equation}\label{p_p_trans_FD}
    k_0\approx-i\frac{\big(\pi/(\mu_p^2 M^2)\big)^{1/3}}{1+\big(\mu_p^2 M^2/\pi\big)^{2/3}}|\spk|.
\end{equation}
As we see, the transverse modes in a degenerate neutron gas have only resonances with $\re k_0=0$. For small $|\spk|$, these resonances can approach rather close to the real axis. The only bounds from below on $|\spk|$ are the requirements that $1/|\spk|$ is much smaller than the size of the domain where a local thermodynamic equilibrium is reached and the estimates \eqref{collision_conds} are satisfied.


\begin{figure}[tp]
\centering
\includegraphics*[width=0.45\linewidth]{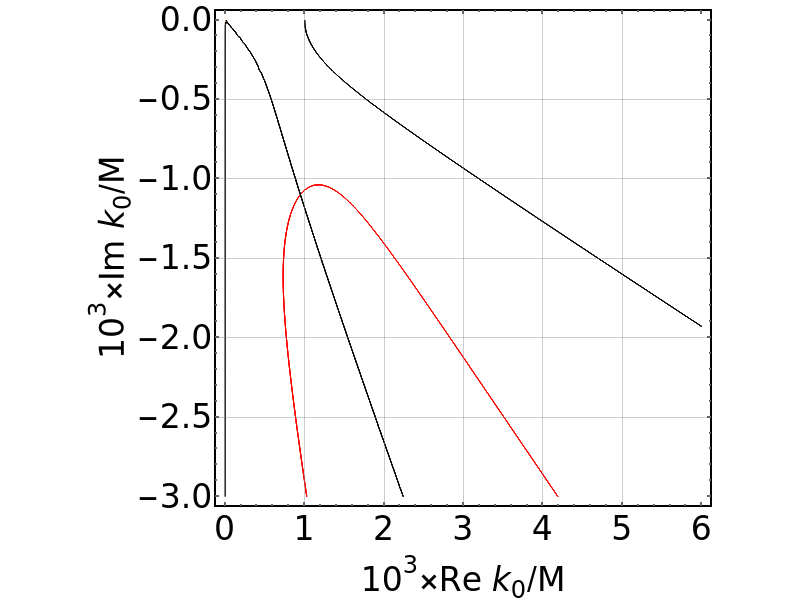}\;
\includegraphics*[width=0.45\linewidth]{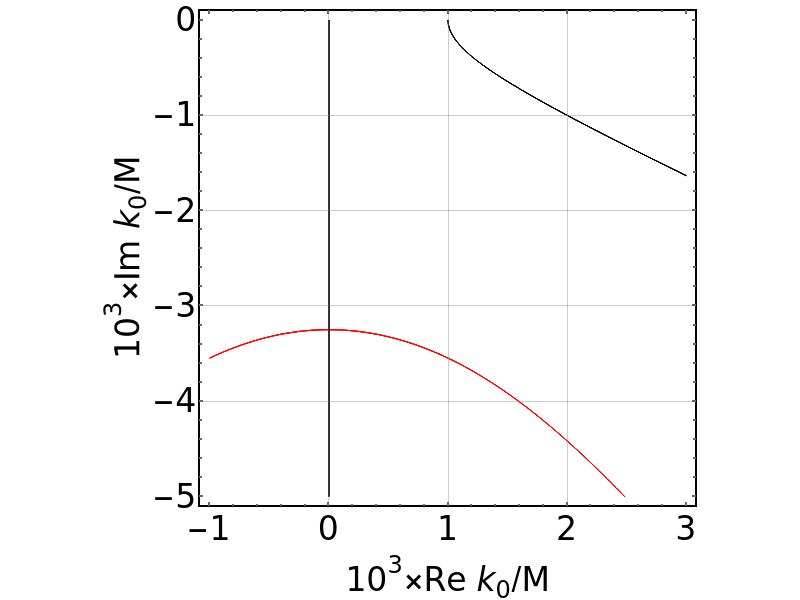}
\caption{{\footnotesize The same as in Fig. \ref{NumSol_MB_1neutr_plots} but for a neutron gas obeying the Fermi-Dirac distribution \eqref{Wign_func_scalar_FD}. The parameters are the same as in Fig. \ref{Phi_chi_FD_stat_plots}. The minimum energy following from \eqref{collision_conds} is $k_0^{min}=7.5\times 10^{-11}M=70$ meV. The plasmon-polariton momentum is $|\spk|=10^{-3}M=940$ keV. Left panel: The solution of Eq. \eqref{longitud_disp_law} for longitudinal plasmon-polaritons. The root of Eq. \eqref{longitud_disp_law} is well approximated by \eqref{p_p_long_FD}. Right panel: The solution of Eq. \eqref{transv_disp_law} for transverse plasmon-polaritons. The root of Eq. \eqref{transv_disp_law} is well approximated by \eqref{p_p_trans_FD}.} }
\label{NumSol_p_p_FD_plots}
\end{figure}


In order to find the dispersion law of longitudinal plasmon-polaritons, it is necessary to know the longitudinal electric susceptibility. For a degenerate neutron gas, it becomes
\begin{equation}
    \chi_\parallel(k)=-\chi_\parallel(0) \bigg\{1+\frac{k^2}{4|\spk|p_F} \Big[\frac{1}{z_+} F_D^{(1)}(\tmu,x_+) -\frac{1}{z_-}F_D^{(1)}(\tmu,x_-)\Big] -\frac{2p_F}{|\spk|} \Big[\frac{1}{z_+} F_D^{(2)}(\tmu,x_+) -\frac{1}{z_-}F_D^{(2)}(\tmu,x_-)\Big] \bigg\},
\end{equation}
where expressions \eqref{F_D_degen0} should be substituted in place of $F_D^{(1,2)}(\tmu,x)$. The equation \eqref{longitud_disp_law} determining the dispersion law of longitudinal modes possesses solutions only in the case $\im z_\pm<0$ and $|z_\pm|\gg1$. Using the asymptotics \eqref{F_D_1_asympt} and
\begin{equation}
    F^{(2)}_D(\tmu,z_\pm)\approx -\frac{15i\pi}{8}z_\pm^5,
\end{equation}
we come to the approximate equation
\begin{equation}
    \Big(\frac{k^2_0}{\spk^2}-1\Big)\frac{k^3_0}{|\spk|^3}=\frac{i\pi}{\mu_p^2M^2},
\end{equation}
whence
\begin{equation}\label{p_p_long_FD}
    k_0\approx  \frac{e^{-3i\pi/10}\big(\pi/(\mu_p^2 M^2)\big)^{1/5}}{1-e^{3i\pi/5}\big(\mu_p^2 M^2/\pi\big)^{2/5}}|\spk|,
\end{equation}
where, recall, $\mu_p^2M^2\approx 4\pi\al$. Thus we see that there are the rapidly decaying longitudinal plasmon-polariton modes with linear dispersion law in a degenerate neutron gas. The numerical solution of Eqs. \eqref{transv_disp_law} and \eqref{longitud_disp_law} in the case of the degenerate neutron gas is presented in Fig. \ref{NumSol_p_p_FD_plots}.

\section{Conclusion}

Let us sum up the results. We studied the behavior of the electromagnetic field in a rarefied unpolarized nonrelativistic neutron gas when the nuclear interaction between neutrons can be neglected. We considered the properties of the electromagnetic field on the space scales much smaller than the typical space scale of variations of the Wigner function of the one-particle density matrix of a neutron gas. The two one-particle density matrices describing the state of a neutron gas were scrutinized: the Gaussian (Maxwell-Boltzmann) one-particle density matrix and the Fermi-Dirac distribution. We investigated the static limit $k_0\rightarrow0$, the long wavelength limit $\spk\rightarrow0$, and the solutions of the free effective Maxwell equations -- plasmon-polaritons. The plasmon-polaritons on the wave packet of a single neutron were also considered.

It turns out that there are the two dimensionless parameters characterizing the magnitude of an electromagnetic response of a neutron gas. They are $\vk=\mu_p^2(x)M/\s^2$ and $\chi_\parallel(0)=\mu_p^2(x)/M$, where $\mu_p^2(x)$ is the density of the neutron anomalous magnetic moment squared and $\s^2$ is the dispersion of momenta in the Wigner function of the one-particle density matrix. The first dimensionless parameter characterizes the behavior of the transverse electromagnetic modes and by the order of magnitude is equal to $1-\mu_\perp^{-1}$, where $\mu_\perp$ is the static magnetic permeability of transverse modes at zero momentum. For the Maxwell-Boltzmann distribution this is an exact equality. The second dimensionless parameter, $\chi_\parallel(0)$, which is much smaller than the first one, is the static longitudinal electric susceptibility at zero momentum. We showed that for a rarefied neutron gas, the quantity $\vk$ cannot be larger than unity even for thermodynamically nonequilibrium states and so the ferromagnetic instability is not realized in this case.

For a nondegenerate neutron gas, we obtained the Green functions \eqref{Green_func_transv0}, \eqref{Green_func_long0} for the static effective Maxwell equations. At the boundary of instability region, $\vk\rightarrow1-0$, the Green function for transverse modes has the form of a Cornell potential of quark-antiquark interaction \cite{Eichten1978} witnessing the presence of anti-screening of the magnetic field in a neutron gas. In the case of a degenerate neutron gas, we found the asymptotics of the Green functions \eqref{Green_func_trans}, \eqref{Green_func_long_asympt} at large distances. It turns out that the analog of Friedel oscillations arises both in the static electric field and in the static magnetic field. These oscillations are subleading on the background of the leading monotonic contribution proportional to $1/r$.

We obtained the approximate expressions for the dispersion laws of longitudinal and transverse plasmon-polaritons in a dilute neutron gas and on a single neutron. We found that there is an infinite number of branches of the plasmon-polariton dispersion law at a given momentum $\spk$ in this case. In spite of the fact that the effect of the anomalous magnetic moments of neutrons in an unpolarized low-density neutron gas is small, the plasmon-polaritons exist for a nondegenerate neutron gas and can possess a sufficient lifetime to be considered as quasiparticles. In a degenerate neutron gas, the plasmon-polaritons appear only as rapidly decaying resonances.

For completeness, we also investigated the plasmon-polaritons on a Gaussian wave packet of a single electron. In the case of isotropic in the momentum space and homogeneous in the coordinate space Wigner functions of the one-particle density matrices, the general expressions for the polarization operators for neutrons \eqref{Pi_par_Pi_perp_gen} and electrons \eqref{Pi_par_Pi_perp_gen_el} are quite similar apart from an interchange of transverse and longitudinal parts and some factors of $2$ and $k^2$. As in the case of a neutron gas, we found that there is an infinite number of branches of the plasmon-polariton dispersion law at a fixed momentum $\spk$. At sufficiently large momenta $\spk$, the plasmon-polaritons corresponding to the branches of the dispersion law with small imaginary part can be considered as quasiparticles. Moreover, even for a single electron there exist stable longitudinal plasmon-polariton modes at certain values of momenta \eqref{long_pp_wo_im1} (see Fig. \ref{NumSol_MB_1electr_plots}).

The plasmon-polaritons in a neutron gas or in a single electron wave packet can be resonantly excited in scattering processes with off-shell photons produced by charged particles, in particular, the bunch instabilities can occur \cite{LandLifPhysKin,AAPSS1975book}. We plan to investigate such processes elsewhere. Among the other directions for a further study we may distinguish the derivation of the photon polarization operator and the description of properties of the respective effective Maxwell equations for a rarefied neutron gas in the magnetic field and in the field of a strong electromagnetic wave, taking into account the spin polarization and the finite size of a neutron gas, and the investigation of relativistic plasmon-polaritons on a single electron wave packet. We leave a detailed study of these problems for future research.

\appendix
\section{Properties of the functions $\Li_\nu(-e^{\tmu})$ and $F^{(\nu)}_D(\tmu,x)$}\label{App_F_D_Expansions}

In evaluating the integrals over the Fermi-Dirac distribution for nonrelativistic particles, there appears the polylogarithm that is equal by definition to
\begin{equation}
    -\Li_\nu(-e^{\tmu}):=\frac{1}{\Ga(\nu)}\int_0^\infty \frac{dx x^{\nu-1}}{e^{x-\tmu}+1},\qquad \re\nu>0,\; |\im\tmu|<\pi.
\end{equation}
For other $\nu\in \mathbb{C}$, the polylogarithm is understood in the sense of an analytical continuation. We provide in this appendix some properties of this function necessary for our investigation. The function $\Li_\nu(-e^{\tmu})$ is an entire function of $\nu$ for $|\im\tmu|<\pi$. Its Mellin transform is
\begin{equation}\label{polylog_mellin_transf}
    \int_0^\infty dx x^{s-1}\Li_\nu (-e^{\tmu-x}) = \Ga(s)\Li_{\nu+s} (-e^{\tmu}),\qquad\re s>0.
\end{equation}
In particular,
\begin{equation}
    \int_0^\infty dx x^{s-1}\Li_\nu (-e^{-x}) = \Ga(s)\Li_{\nu+s} (-1)=-\Ga(s)\eta(\nu+s),
\end{equation}
where
\begin{equation}
    \eta(\nu)=(1-2^{1-\nu})\zeta(\nu),
\end{equation}
and $\zeta(\nu)$ is the zeta function. Recall that $\eta(\nu)$ is an entire function of $\nu\in \mathbb{C}$. The Mellin representations of the polylogarithm read
\begin{subequations}
\begin{align}
    -\Li_\nu(-e^{\tmu})&=\int_{(0,1)} \frac{ds}{2\pi i}\frac{\pi}{\sin(\pi s)} \frac{e^{s\tmu}}{s^\nu},&\qquad |\im{\tmu}|&<\pi,\label{polylog_mellin_repr1}\\
    \Li_\nu(-e^{\tmu-x})&=\int_{(0,\infty)} \frac{ds}{2\pi i} x^{-s} \Ga(s)\Li_{\nu+s} (-e^{\tmu}),&\qquad |\arg x|&<\pi/2,\label{polylog_mellin_repr2}
\end{align}
\end{subequations}
where $\int_{(a,b)}ds$ means that the integration is carried out along the contour parallel to the imaginary axis lying in the strip $\re s\in(a,b)$ and going from bottom to top. Stretching the integration variable, $s\rightarrow s/\tmu$, in the representation \eqref{polylog_mellin_repr1} and using the expansion
\begin{equation}
    \frac{\pi s}{\sin(\pi s)}=\sum_{k=0}^\infty 2\eta(2k)s^{2k},
\end{equation}
we obtain the asymptotic expansion,
\begin{equation}\label{polylog_large_mu}
    -\Li_{\nu}(-e^{\tmu})\simeq \sum_{k=0}^\infty \frac{2\eta(2k)}{\Ga(\nu+1-2k)} \tmu^{\nu-2k},
\end{equation}
up to exponentially suppressed terms at $\tmu\rightarrow+\infty$. Moving the integration contour to the left with account for the singularities of the integrand in the representation \eqref{polylog_mellin_repr2}, we deduce the expansion for small $\tmu$:
\begin{equation}\label{polylog_expans_zero}
    -\Li_\nu(-e^{\tmu})=\sum_{k=0}^\infty \eta(\nu-k)\frac{\tmu^k}{k!},\qquad |\tmu|<\pi.
\end{equation}
In order to obtain this expansion, one needs to put $\tmu\rightarrow0$ and $x\rightarrow-\tmu$ in \eqref{polylog_mellin_repr2}. Shifting the integration contour to the left with account for the singularities of the integrand in the representation \eqref{polylog_mellin_repr1}, we arrive at the standard series representation of the polylogarithm
\begin{equation}
    -\Li_\nu(-e^{\tmu})=\sum_{k=1}^\infty \frac{(-1)^{k-1}}{k^\nu}e^{k\tmu},\qquad \re\tmu<0.
\end{equation}
Consequently, we have for $\tmu\ll-1$,
\begin{equation}\label{polylog_negative_tmu}
    -\Li_\nu(-e^{\tmu}) \approx e^{\tmu}.
\end{equation}
Henceforth, we suppose that $\tmu\in \mathbb{R}$.

The expression for the particle number density of neutrons \eqref{dens_neutr_gas} defines the chemical potential in terms of the density of a neutron gas. Employing the approximate equality \eqref{polylog_negative_tmu}, the Fermi-Dirac distribution turns into the Maxwell-Boltzmann one for \cite{LandLifStatPhysP1}
\begin{equation}\label{nondeg_gas_cond}
    \be\ll\be_F,\qquad \be_F:=2M (3\pi^2\rho(x))^{-2/3}.
\end{equation}
This condition is equivalent to
\begin{equation}
    \rho(x)\ll \frac14\Big(\frac{2M}{\pi\beta}\Big)^{3/2}.
\end{equation}
The region $\be\mu\gg1$ corresponds to a degenerate Fermi gas, whereas $\be\mu\approx0$ is an intermediate regime arising for a neutron gas with the particle number density
\begin{equation}
    \rho(x)\approx \frac{\eta(3/2)}{4}\Big(\frac{2M}{\pi\beta}\Big)^{3/2}\approx 0.1 \Big(\frac{M}{\beta}\Big)^{3/2}.
\end{equation}
Here we have used the expansion of the polylogarithm \eqref{polylog_expans_zero} for a small dimensionless chemical potential.

In evaluating the polarization operator for a rarefied neutron gas in a thermodynamic equilibrium, the following function appears
\begin{equation}\label{F_D_func}
    F^{(\nu)}_D(\tmu,x):=\frac{1}{\Li_{\nu+1/2}(-e^{\tmu})}\frac{x}{\sqrt{\pi}}\int_{-\infty}^\infty \frac{dz}{z-x} \Li_{\nu}(-e^{\tmu-z^2})= \frac{1}{\Li_{\nu+1/2}(-e^{\tmu})} \frac{x^2}{\sqrt{\pi}} \int_{-\infty}^\infty \frac{dz}{z^2-x^2} \Li_\nu(-e^{\tmu-z^2}),
\end{equation}
where $\im x>0$ and it is assumed that the principal branch of the polylogarithm is chosen. The function $F^{(\nu)}_D(\tmu,x)$ is analytical for $\im x>0$ and real-valued for $x=i y$, $y>0$, and $\nu\in \mathbb{R}$. Therefore, the symmetry property holds
\begin{equation}\label{F_D_symm}
    F^{(\nu)*}_D(\tmu,x)=F^{(\nu)}_D(\tmu,-x^*),\qquad\nu\in \mathbb{R}.
\end{equation}
For real $\nu$ and $x$, we obtain
\begin{equation}
    \re F^{(\nu)}_D(\tmu,x)=\re F^{(\nu)}_D(\tmu,-x),\qquad \im F^{(\nu)}_D(\tmu,x)=-\im F^{(\nu)}_D(\tmu,-x).
\end{equation}
For $\im x\leqslant 0$, the function $F^{(\nu)}_D(\tmu,x)$ is understood in the sense of an analytical continuation.

Let us investigate some properties of the function $F^{(\nu)}_D(\tmu,x)$ and obtain its asymptotic expansions in different regions of its arguments supposing that $\tmu\in \mathbb{R}$ and $x\in \mathbb{C}$. It is evident that for $\tmu\ll-1$, which corresponds to a nondegenerate Fermi gas,
\begin{equation}\label{F_D_nondeg}
    F^{(\nu)}_D(\tmu,x)\approx F(x),
\end{equation}
and expressions \eqref{Pi_perp_FD} and \eqref{chi_parallel_FD} go into \eqref{Pi_perp_Max} and \eqref{Pi_parallel_Max}, respectively, where one should put $\s=\sqrt{M/\beta}$.

The function \eqref{F_D_func} has the integral representations
\begin{subequations}\label{F_D_represent}
\begin{align}
    F^{(\nu)}_D(\tmu,x)&=\frac{1}{\Li_{\nu+1/2}(-e^{\tmu})}\Big[ \frac{x}{\sqrt{\pi}}\dashint_{-\infty}^\infty \frac{dz}{z-x} \Li_\nu(-e^{\tmu-z^2}) +i\sqrt{\pi}x \Li_\nu(-e^{\tmu-x^2})\Big],&\qquad \im x&=0,\\
    F^{(\nu)}_D(\tmu,x)&=\frac{1}{\Li_{\nu+1/2}(-e^{\tmu})}\Big[ \frac{x}{\sqrt{\pi}}\int_{-\infty}^\infty \frac{dz}{z-x} \Li_\nu(-e^{\tmu-z^2}) +2i\sqrt{\pi}x \Li_\nu(-e^{\tmu-x^2})\Big],&\qquad \im x&<0.\label{F_D_Im_x_neg}
\end{align}
\end{subequations}
The following symmetry properties are fulfilled
\begin{equation}\label{F_D_symm_props}
    F^{(\nu)}_D(\tmu,-x)=F^{(\nu)}_D(\tmu,x) -2i\sqrt{\pi}x \frac{\Li_\nu(-e^{\tmu-x^2})}{\Li_{\nu+1/2}(-e^{\tmu})},\qquad F^{(\nu)*}_D(\tmu,x)=F^{(\nu)}_D(\tmu,x^*) -2i\sqrt{\pi}x^* \frac{\Li_\nu(-e^{\tmu-(x^*)^2})}{\Li_{3/2}(-e^{\tmu})}.
\end{equation}
It follows from the representations \eqref{F_D_represent} that the branch of the function $F_D(\tmu,x)$ we consider has the branch points in the lower half-plane, $\im x<0$, at
\begin{equation}\label{branch_points}
    x=-\sqrt{\tmu+i\pi(2n+1)},\qquad n\in \mathbb{Z},
\end{equation}
where such a branch of the square root is chosen that takes the values in the upper half-plane.

Let $\im x>0$. Then substituting the Mellin representation of the polylogarithm \eqref{polylog_mellin_repr2} into \eqref{F_D_func} and evaluating the integral
\begin{equation}
    \int_{-\infty}^\infty \frac{dzz^{-2s}}{z-x}=\frac{i\pi e^{i\pi s}}{\cos(\pi s)}x^{-2s},\qquad \re s\in(0,1/2),
\end{equation}
we have
\begin{equation}\label{F_D_Mellin}
    F_D^{(\nu)}(\tmu,x)=\frac{i\sqrt{\pi} x}{\Li_{\nu+1/2}(-e^{\tmu})} \int_{(0,1/2)} \frac{ds}{2\pi i} x^{-2s} \frac{e^{i\pi s}\Ga(s)}{\cos(\pi s)}\Li_{\nu+s}(-e^{\tmu}),\qquad |\arg x^2|<\pi/2.
\end{equation}
Shifting the integration contour in this integral to the right with account for the singularities of the integrand, we come to
\begin{equation}\label{F_D_asympt}
    F^{(\nu)}_D(\tmu,x) \simeq -\sum_{n=0}^\infty \frac{ \Gamma(1/2+n) \Li_{\nu+1/2+n}(-e^{\tmu} ) }{ \Gamma(1/2) \Li_{\nu+1/2}(-e^{\tmu} ) } x^{-2n}\approx -1-\frac{\Li_{\nu+3/2}(-e^{\tmu} )}{\Li_{\nu+1/2}(-e^{\tmu} )}\frac{1}{2x^2},\qquad \im x>0.
\end{equation}
This expansion is valid for $|x|^2\gg\max(1,\tmu)$ and is a generalization of the asymptotic expansion \eqref{F_asympt}. The expansions of $F^{(\nu)}_D(\tmu,x)$ for $\im x\leqslant0$ can be obtained with the aid of the first symmetry property in \eqref{F_D_symm_props}.

Moving the integration contour in the integral \eqref{F_D_Mellin} to the left with account for the singularities of the integrand, we deduce
\begin{equation}\label{F_D_expans_zero}
    F^{(\nu)}_D(\tmu,x)=i\sqrt{\pi}\sum_{n=0}^\infty \frac{(-1)^n}{n!} \frac{\Li_{\nu-n}(-e^{\tmu})}{\Li_{\nu+1/2}(-e^{\tmu})}x^{2n+1} -\sum_{n=0}^\infty \frac{\sqrt{\pi}(-1)^n}{\Ga(n+3/2)} \frac{\Li_{\nu-1/2-n}(-e^{\tmu})}{\Li_{\nu+1/2}(-e^{\tmu})}x^{2n+2}.
\end{equation}
This expansion converges in the vicinity of the point $x=0$ and is a generalization of the expansion \eqref{F_small_arg}.

The representation \eqref{F_D_Mellin} can be employed to derive the expansion of $F_D^{(\nu)}(\tmu,x)$ for large $\tmu$, i.e., for a degenerate Fermi gas. Substituting the expansion \eqref{polylog_large_mu} into \eqref{F_D_Mellin}, we have
\begin{equation}
    F_D^{(\nu)}(\tmu,x) \simeq -i\sqrt{\pi} \sum_{n=0}^\infty \frac{2\eta(2n)x\tmu^{\nu-2n}}{\Li_{\nu+1/2}(-e^{\tmu})} \int_{(0,1/2)} \frac{ds}{2\pi i} \Big(\frac{x^2}{\tmu}\Big)^{-2s} \frac{e^{i\pi s}\Ga(s)}{\cos(\pi s)\Ga(\nu+1-2n+s)},
\end{equation}
up to exponentially suppressed terms at $\tmu\rightarrow+\infty$. The appearing integral over $s$ is expressed through the hypergeometric functions. As a result, we obtain the asymptotic expansion
\begin{equation}\label{F_D_large_tmu}
    F_D^{(\nu)}(\tmu,x) \simeq - \sum_{n=0}^\infty \frac{2\eta(2n)}{\Li_{\nu+1/2}(-e^{\tmu})} \tmu^{\nu-2n+1/2} z\Big[i\sqrt{\pi} \frac{(1-z^2)^{\nu-2n}}{\Ga(\nu-2n+1)} -2z\frac{F(1,1/2+2n-\nu;3/2;z^2)}{\Ga(\nu-2n+1/2)}\Big],
\end{equation}
where $z:=x/\sqrt{\tmu}$ and $\Li_{\nu+1/2}(-e^{\tmu})$ in the denominator has to be expanded as in formula \eqref{polylog_large_mu}. In virtue of the uniqueness of analytical continuation, the expansion \eqref{F_D_large_tmu} is valid for any $x\in \mathbb{C}$. We need the explicit expressions
\begin{equation}\label{F_D_degen0}
\begin{split}
    F^{(1)}_D(\tmu,x)&\approx \frac{3}{2} z \Big[-z+\frac{1-z^2}{2} \big(i\pi+\ln\frac{1-z}{1+z} \big)\Big],\\
    F^{(2)}_D(\tmu,x)&\approx \frac{15}{8} z \Big[-\frac{5}{3}z+z^3+\frac{(1-z^2)^2}{2} \big(i\pi+\ln\frac{1-z}{1+z} \big)\Big],
\end{split}
\end{equation}
where only the leading contributions at $\tmu\rightarrow+\infty$ are retained and the principal branch of the logarithm is taken. Analogously, substituting the expansion \eqref{polylog_expans_zero} into \eqref{F_D_Mellin}, one can find the expansion of $F^{(\nu)}_D(\tmu,x)$ in ascending powers of $\tmu$. However, as we do not need such an expansion in our study, we do not write it here.

\section{Justification of the small $q$ approximation}\label{Small_q_Approx_App}

We show in this appendix that under the conditions \eqref{k_nonrel}, \eqref{q_less_sigma}, and \eqref{large_k_conds} one can neglect the dependence of the tensor $G_3^{\mu\nu}$ in the numerator and of $p_0$ in the denominator of the integrand in \eqref{polar_oper0} on $q$. Indeed, in the nonrelativistic approximation,
\begin{equation}
    p_0(\spp_c+\spq/2)p_0(\spp_c-\spq/2)\approx\Big(M+\frac{\spp_c^2}{2M}\Big)^2 - \frac{(\spp_c\spq)^2}{4M^2}.
\end{equation}
The term depending on $\spq$ is of order $\s^4/M^4$ in comparison with the leading term and it can be omitted. As for the tensor $G^{\mu\nu}_3$, it has the form (see formulas (46), (50) of \cite{ComptNeutr})
\begin{equation}
    G^{\mu\nu}=\frac{P^{\mu\nu}(p_c,k,q)}{((p_c-k)^2-M^2) ((p_c+k)^2-M^2)\sqrt{(p_0(\spp+\spq/2)+M) (p_0(\spp-\spq/2)+M)}},
\end{equation}
where
\begin{equation}
    p^c_0=\big(p_0(\spp+\spq/2)+p_0(\spp-\spq/2)\big)/2,
\end{equation}
and $P^{\mu\nu}(p_c,k,q)$ is a polynomial in the components of the momenta $p^\mu_c$, $k^\mu$, and $q^\mu$. The expression under square root sign is written as
\begin{equation}
    \big(p_0(\spp+\spq/2)+M\big) \big(p_0(\spp-\spq/2)+M\big)\approx\Big(2M+\frac{\spp_c^2}{2M}\Big)^2 - \frac{(\spp_c\spq)^2}{4M^2},
\end{equation}
and, therefore, its dependence on $\spq$ can be neglected. Furthermore,
\begin{equation}
    (p_c \pm k)^2-M^2\approx\pm 2\big(p_0(\spp_c)k_0-\spp_c\spk\big) +k^2 \pm k_0 \frac{\spq^2}{4M}-\frac{(\spp_c\spq)^2}{M^2} -\spq^2.
\end{equation}
As is seen, under the conditions \eqref{large_k_conds} the terms depending on $\spq$ can be neglected in comparison with the contribution of $k^2$. As long as
\begin{equation}
    |q_0|\sim\frac{\s^2}{M},
\end{equation}
and the estimate \eqref{q_less_sigma} is valid, the first three conditions in \eqref{small_q_conds1} are fulfilled. Consequently, one can neglect the terms depending on $\spq$ in the polynomial $P^{\mu\nu}(p_c,k,q)$ in comparison with the terms depending only on $p_c$ and $k$. By the same reasons, one can discard the terms depending on $\spp_c$ in $P^{\mu\nu}(p_c,k,q)$. However, we keep these small contributions to $P^{\mu\nu}(p_c,k,q)$ for uniformity of notation.

\section{Plasmon-polaritons on a single electron}\label{Plasm-Pol_on_Singl_El_App}

In the paper \cite{AKS2025}, the plasmon-polaritons on the wave packet of a single electron were investigated in the approximation where the electron wave packet is very narrow in the momentum space. In this appendix, we consider a more realistic model of the electron wave packet. Namely, we describe the properties of short wavelength plasmon-polaritons on a single electron in the case when its state is not spin polarized and is described by the density matrix with the scalar Wigner function \eqref{Wign_func_scalar_MB}. We suppose that this state is pure, i.e., $\s_x=1/(2\s)$, and the conditions of the short wavelength and nonrelativistic approximations  \eqref{short_wave_appr}, \eqref{k_nonrel}, \eqref{small_q_sigma_x}, \eqref{nonrel_appr_2} are satisfied, where $M$ is the electron mass. For uniformity of notation, we shall denote the electron mass as $M$ in this appendix. We shall neglect the vacuum contribution to the photon polarization operator, which is justified under the approximations we use \cite{AKS2025}. In addition to being of independent interest, the analysis of properties of plasmon-polaritons on a single electron allows us to compare them with the properties of plasmon-polaritons on a single neutron.

The general expression for the photon polarization operator in the presence of an inhomogeneous electron gas and, in particular, in the presence of a single electron, prepared in an arbitrary quantum state was derived in formula (56) of \cite{AKS2025}. For a spin unpolarized state of the electron, the Weyl symbol of this polarization operator with the above approximations is given by
\begin{equation}
    \Pi^{\mu\nu}(x,k)=e^2\int\frac{d\spp}{p_0} \rho(x,\spp)\frac{(kp)^2\eta^{\mu\nu} -(kp)k^{(\mu}p^{\nu)} +k^2 p^\mu p^\nu}{(kp)^2-k^4/4}.
\end{equation}
For a uniform in space Wigner function of the one-particle density matrix, this expression coincides with formula (8.3.14) of the book \cite{Melrose2008} (see also \cite{Braaten1993}). It follows from this expression that (cf. \eqref{Pi_par_Pi_perp_gen})
\begin{equation}\label{Pi_par_Pi_perp_gen_el}
    \Pi_\parallel(k)=e^2M^2 k^2\int \frac{d\spp}{p_0}\rho(x,\spp)\frac{1+\spp_\perp^2/M^2}{(kp)^2-k^4/4},\qquad
    \Pi_\perp(k)=e^2 \int \frac{d\spp}{p_0}\rho(x,\spp)\Big[1 -\frac{k^2}{2} \frac{\spp_\perp^2 -k^2/2}{(kp)^2-k^4/4} \Big].
\end{equation}
These expressions coincide with the expressions presented in formula (9.1.7) of \cite{Melrose2008}. The quantity $\spp_\perp$ was introduced in formula \eqref{Pi_par_Pi_perp_gen}. Let us introduce, for brevity, the notation
\begin{equation}
    \tilde{\Phi}(k):=\frac{F(x_+)}{\alpha_+}  -\frac{F(x_-)}{\alpha_-},
\end{equation}
where $x_\pm$ were defined in \eqref{x_pm_defn}. Then, in the nonrelativistic approximation,
\begin{equation}
    \Pi_\parallel(k)\approx M^2\omega_p^2\tilde{\Phi}(k),\qquad \Pi_\perp(k)\approx\omega_p^2\big[1 +\big(k^2/4-\s^2\big)\tilde{\Phi}(k)\big],
\end{equation}
where $\omega_p^2=e^2\rho(x)/M$ is the plasma frequency squared. The similar expressions for the Fermi-Dirac distribution look as (cf.  \eqref{Pi_perp_FD} and \eqref{chi_parallel_FD})
\begin{equation}
\begin{split}
    \Pi_\parallel(k)&\approx M^2\omega_p^2\Big[\frac{F_D^{(1)}(x_+)}{\alpha_+}  -\frac{F_D^{(1)}(x_-)}{\alpha_-}\Big],\\ \Pi_\perp(k)&\approx \omega_p^2\Big\{1+\frac{k^2}{4}\Big[\frac{F_D^{(1)}(x_+)}{\alpha_+}  -\frac{F_D^{(1)}(x_-)}{\alpha_-}\Big] -\s^2\frac{\Li_{5/2}(-e^{\tmu})}{\Li_{3/2}(-e^{\tmu})} \Big[\frac{F_D^{(2)}(x_+)}{\alpha_+}  -\frac{F_D^{(2)}(x_-)}{\alpha_-}\Big] \Big\}.
\end{split}
\end{equation}

\begin{figure}[!tp]
\centering
\includegraphics*[width=0.33\linewidth]{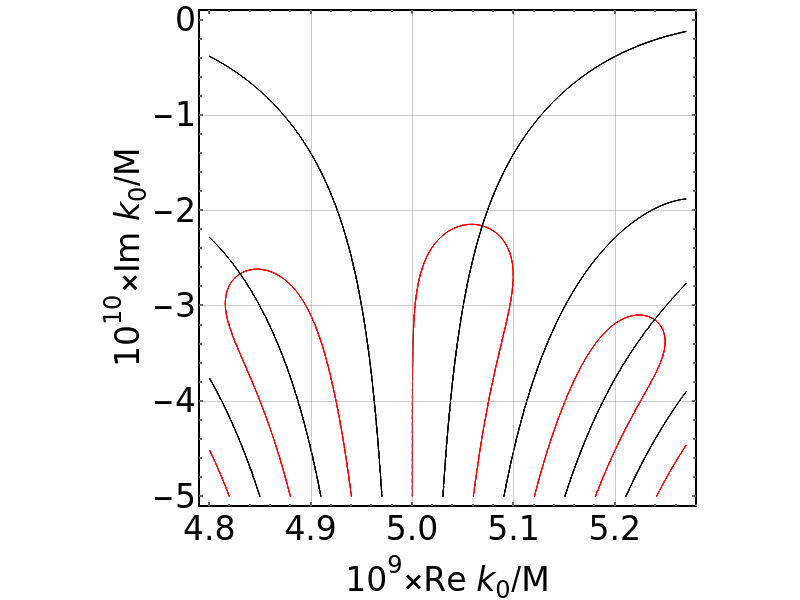}\;
\includegraphics*[width=0.33\linewidth]{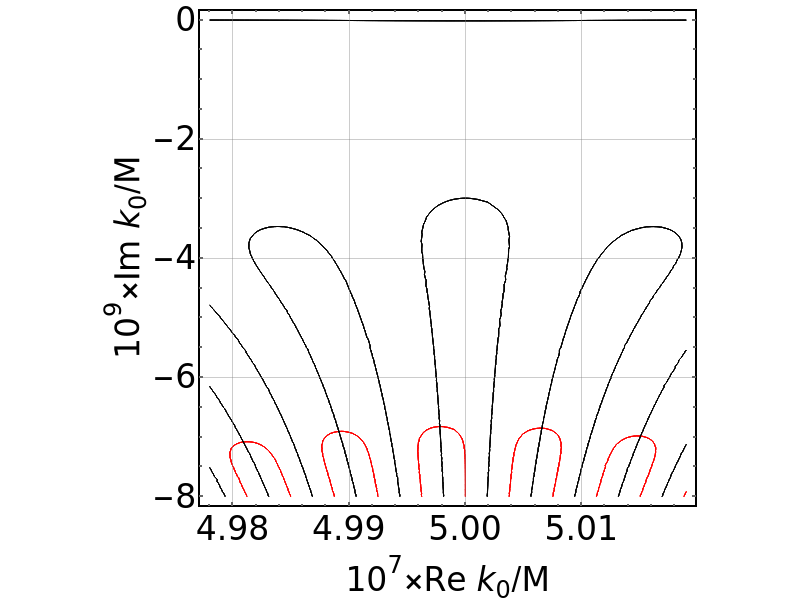}\\
\includegraphics*[width=0.33\linewidth]{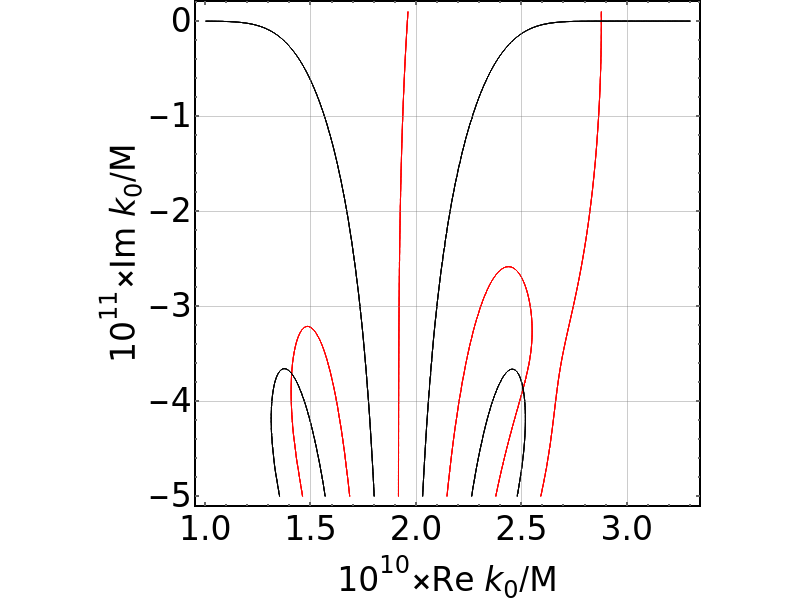}\;
\includegraphics*[width=0.33\linewidth]{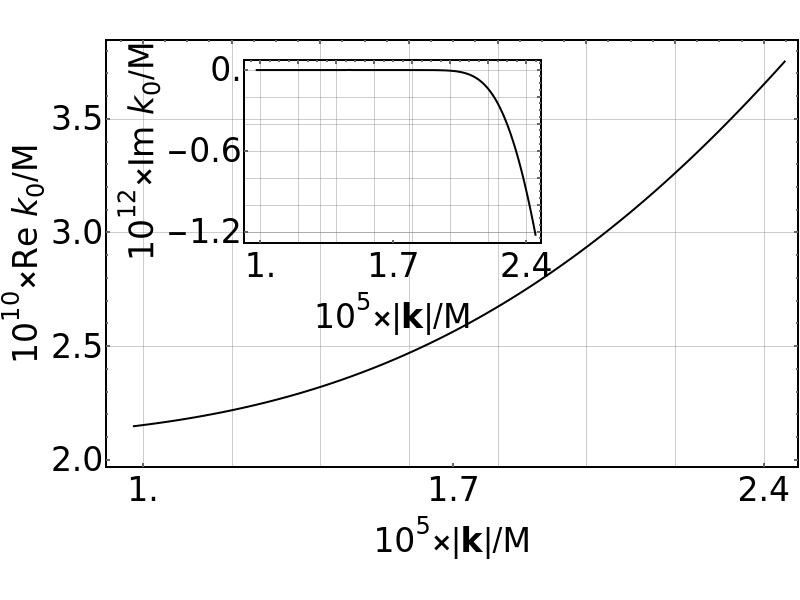}
\caption{{\footnotesize The same as in Fig. \ref{NumSol_MB_1neutr_plots} but for the Gaussian wave packet \eqref{Wign_func_scalar_MB} of a single electron with $\s=9.78\times 10^{-7} M=0.5$ eV and $\s_x=1/(2\s)=0.20$ $\mu$m. Top left panel: The solution of Eq. \eqref{disp_law_longit_el0} for longitudinal plasmon-polaritons with momentum $|\spk|=10^{-4}M=51.1$ eV. The roots of Eq. \eqref{disp_law_longit_el0} are well approximated by \eqref{disp_law_longit_el}. It is seen that the relative magnitude of the imaginary part of the energy of plasmon-polaritons is much less than its real part and so these plasmon-polaritons can be interpreted as unstable quasiparticles. Top right panel: The solution of Eq. \eqref{disp_law_trans_el0} for transverse plasmon-polaritons with momentum $|\spk|=10^{-3}M=511$ eV. The roots of Eq. \eqref{disp_law_trans_el0} are well approximated by \eqref{disp_law_trans_el_appr}. Bottom left panel: The solution of Eq. \eqref{disp_law_longit_el0} for longitudinal plasmon-polaritons with momentum $|\spk|=1.95\times 10^{-5}M=10$ eV. It is seen that one of the roots of Eq. \eqref{disp_law_longit_el0} is almost real and is well approximated by \eqref{disp_law_longit_el1}. Bottom right panel: The dependence of the real and imaginary parts of this root on $|\spk|$. The real part is well approximated by \eqref{disp_law_longit_el1}. } }
\label{NumSol_MB_1electr_plots}
\end{figure}

Let us obtain the approximate expression for the dispersion law of longitudinal plasmon-polaritons on the wave packet of a single electron prepared in the state with the scalar Wigner function \eqref{Wign_func_scalar_MB}. The equation \eqref{disp_law_longit0} is written as
\begin{equation}\label{disp_law_longit_el0}
    \frac{M^2\omega_p^2}{k^2}\tilde{\Phi}(k)=1.
\end{equation}
In the case \eqref{case1_long_plasm_pol}, for $\im k_0<0$ we deduce
\begin{equation}\label{tPhi_appr}
    \tilde{\Phi}(k)\approx\frac{\sqrt{2\pi}i}{|\spk|\s}e^{-\frac{M^2(\de k_0)^2}{2\spk^2\s^2}},
\end{equation}
where $\de k_0=k_0-\spk^2/(2M)$. Solving Eq. \eqref{disp_law_longit_el0}, we arrive at
\begin{equation}\label{disp_law_longit_el}
    \de k_0=-i\frac{|\spk|\s}{M} \sqrt{2\ln\frac{i|\spk|^3\s}{\sqrt{2\pi}M^2\omega_p^2}}.
\end{equation}
Just as in the case of a nondegenerate neutron gas, the dispersion law of longitudinal plasmon-polaritons on the wave packet of a single electron is close to the dispersion law of plasmons $k_0\approx\spk^2/(2M)$. Moreover, there is an infinite number of branches of this dispersion law corresponding to a different choice of the branches of the logarithm in \eqref{disp_law_longit_el}. The relative magnitude of the imaginary correction to the real part of the energy of plasmon-polariton is small in the region of parameters we consider (see Fig. \ref{NumSol_MB_1electr_plots}).

For
\begin{equation}\label{long_pp_wo_im}
    \frac{|\spk|^3\s}{\sqrt{2\pi}M^2\omega_p^2}\lesssim 1,
\end{equation}
the imaginary part of one of the roots of Eq. \eqref{disp_law_longit_el0} becomes negligibly small, and we come to the regime investigated in \cite{AKS2025}. Keeping in mind that
\begin{equation}\label{omega_p}
    \frac{\omega_p^2}{M^2}=8\sqrt{\frac{2}{\pi}}\al\frac{\s^3}{M^3},
\end{equation}
the condition \eqref{long_pp_wo_im} holds when
\begin{equation}\label{long_pp_wo_im1}
    1\ll\frac{|\spk|}{\s}\lesssim \Big(16\al\frac{M}{\s}\Big)^{1/3}.
\end{equation}
In this case, equation \eqref{disp_law_longit_el0} possesses an approximate solution
\begin{equation}\label{disp_law_longit_el1}
    k_0\approx \frac{1}{2M}\sqrt{\spk^4+4 M^2\omega_p^2}.
\end{equation}
Aside from the real root \eqref{disp_law_longit_el1}, equation \eqref{disp_law_longit_el0} has an infinite number of roots of the form \eqref{disp_law_longit_el} (see Fig. \ref{NumSol_MB_1electr_plots}). They describe decaying resonances.


Now we turn to the properties of transverse plasmon-polaritons. The equation \eqref{disp_law_transv0} determining their dispersion law can be cast into the form
\begin{equation}\label{disp_law_trans_el0}
    1+(k^2/4-\s^2)\tilde{\Phi}(k) =k^2/\omega_p^2.
\end{equation}
In the case \eqref{case1_long_plasm_pol} for $\im k_0<0$, it follows from \eqref{tPhi_appr} that approximately
\begin{equation}
    1-\frac{\sqrt{2\pi} i|\spk|}{4\s}e^{-\frac{M^2(\de k_0)^2}{2\spk^2\s^2}} =-\frac{\spk^2}{\omega_p^2},
\end{equation}
where it has been assumed that $|k_0|\ll|\spk|$. Hence,
\begin{equation}\label{disp_law_trans_el_appr}
    \de k_0= -i\frac{|\spk|\s}{M}\sqrt{2\ln\Big[\frac{4\s}{\sqrt{2\pi}i|\spk|}\Big(1+\frac{\spk^2}{\omega_p^2}\Big)\Big]}\approx -i\frac{|\spk|\s}{M}\sqrt{2\ln \frac{4\s|\spk|}{\sqrt{2\pi}i\omega_p^2}},
\end{equation}
where, in the last approximate equality, it has been taken into account that $\spk^2/\omega_p^2\gg1$ for the plasma frequency of a Gaussian wave packet of a single electron \eqref{omega_p}. As we see, there is an infinite number of branches of the dispersion law of transverse modes at a fixed momentum $\spk$ corresponding to a different choice of the branches of the logarithm. Since the estimate is fulfilled,
\begin{equation}
    \frac{4\s|\spk|}{\sqrt{2\pi}\omega_p^2}\gg1,
\end{equation}
the energy of transverse modes on the wave packet of a single electron always has an imaginary part in contrast to the longitudinal plasmon-polaritons. Nevertheless, its relative magnitude is much smaller than the real part of the energy for sufficiently large momenta $\spk$ (see Fig. \ref{NumSol_MB_1electr_plots}).

\section{Asymptotics of the polarization operator for $|\spk|\rightarrow0$}\label{Asympt_k_to_0_App}

In this appendix, we obtain the expression for the photon polarization operator in the long wavelength limit $|\spk|\rightarrow0$ in the presence of a dilute neutron gas in the nonrelativistic approximation. The scalar Wigner functions for the one-particle density matrix of the neutron gas are supposed to be \eqref{Wign_func_scalar_MB} or \eqref{Wign_func_scalar_FD}. Using the asymptotics \eqref{F_asympt_2}, \eqref{F_D_asympt}, and the relation
\begin{equation}
    \al_-^{-1} -\al_+^{-1}\approx1/M^2,
\end{equation}
we deduce from \eqref{Pi_perp_Max}, \eqref{Pi_parallel_Max}, \eqref{Pi_perp_FD}, \eqref{chi_parallel_FD} in both cases that
\begin{equation}\label{polar_oper_long_wave}
    \Pi_\perp(k)\approx \Pi_\parallel(k)\approx k_0^2\mu_p^2(x)/M=k_0^2\chi_\parallel(0),
\end{equation}
where the nonrelativistic approximation has been employed. In particular, as follows from the last condition in \eqref{nonrel_appr_2}, expressions \eqref{polar_oper_long_wave} are valid for $|k_0|$ much larger that the average kinetic energy of neutrons in the gas. As a result,
\begin{equation}
    \Pi^{00}\approx -\spk^2\chi_\parallel(0),\qquad \Pi^{0i}\approx -k_0k^i\chi_\parallel(0),\qquad \Pi^{ij}\approx - \de_{ij}k_0^2   \chi_\parallel(0).
\end{equation}
This polarization operator describes an electromagnetic response of a uniform neutron gas under the action of an external weakly inhomogeneous electromagnetic field that varies with time.

\section{Plasmon-polaritons in a Fermi-Dirac nondegenerate neutron gas}\label{Plasm-Polar_NonDeg_Fermi_App}

It has been noted in Secs. \ref{Long_Plams_Polar} and \ref{Trans_Plasm_Polar} that expressions \eqref{y_longitud} and \eqref{y_transv} for the corrections to the energy of longitudinal and transverse plasmon-polaritons are inapplicable in the case of an equilibrium nondegenerate neutron gas. Indeed, in solving Eqs. \eqref{longitud_disp_law}, \eqref{transv_disp_law} in the domain of parameters \eqref{case1_long_plasm_pol}, the asymptotic representation \eqref{F_asympt_3} is used for the function $F(x_+)$, where $x_+$ is defined in \eqref{x_pm_defn}. This asymptotics is valid for a Gaussian one-particle density matrix. For this representation to be true for the Fermi-Dirac distribution in the nondegenerate case, $\tmu\ll-1$, the fulfillment of the condition is necessary
\begin{equation}\label{MB_applic_cond}
    |x^2_+|<-\tmu.
\end{equation}
In this case, the approximate equality \eqref{F_D_nondeg} holds, where $x$ should be replaced by $x_+$. Since for an equilibrium nondegenerate Fermi gas
\begin{equation}
    e^{\tmu}\approx\sqrt{2}\pi^{3/2}\rho(x)/\s^3,
\end{equation}
the condition \eqref{MB_applic_cond} with account for expressions \eqref{longitud_disp_law_case1} and \eqref{transv_disp_law_case1} becomes
\begin{equation}
    \al\frac{|\spk|\s^2}{M^3}\geqslant1,\qquad \al\frac{\s^2}{M |\spk|}\geqslant1,
\end{equation}
for the longitudinal and transverse plasmon-polaritons, respectively. It is evident that these inequalities are violated in the region of parameters \eqref{case1_long_plasm_pol}.

However, it would be strange if the plasmon-polaritons existed for the Maxwell-Boltzmann distribution and did not exist for the Fermi-Dirac one despite the fact that for a nondegenerate Fermi gas, $\tmu\ll-1$, these distributions are virtually indistinguishable. Let us find the approximate expressions for the dispersion laws of longitudinal and transverse plasmon-polaritons in an equilibrium nondegenerate neutron gas in the domain of parameters \eqref{case1_long_plasm_pol}. We assume that $|k_0|\ll|\spk|$. Then from \eqref{longitud_disp_law}, \eqref{chi_parallel_FD}, and \eqref{F_D_Im_x_neg} the approximate equation for the longitudinal modes follows
\begin{equation}\label{longitud_disp_law_case1_nondeg}
    e^{-\tmu}\ln(1+e^{\tmu-x_+^2})\approx\frac{4iM\s}{\sqrt{2\pi}\mu_p^2(x)|\spk|}.
\end{equation}
It is clear that for $\re(\tmu-x_+^2)\ll-1$ this equation turns into \eqref{longitud_disp_law_case1}. However, as it has been discussed above, this estimate is not fulfilled in the case at hand and we need to solve Eq. \eqref{longitud_disp_law_case1_nondeg} exactly. This equation can be written as
\begin{equation}\label{long_eq1}
    \ln(1+e^{\tmu-x_+^2})=\frac{4\pi iM}{\mu_p^2 |\spk|\s^2}=:iL.
\end{equation}
It is easy to see that $L\gg 1$. The equality \eqref{long_eq1} cannot be valid for the principal branch of the logarithm. Nevertheless, it can be satisfied if one passes to the $k$-th branch of the logarithm for sufficiently large $k$. On the $k$-th sheet of the logarithm, equation \eqref{long_eq1} becomes
\begin{equation}\label{long_eq2}
    \ln(1+e^{\tmu-x_+^2})=i(L-2\pi k),
\end{equation}
where $\ln z$ denotes the principal branch of the logarithm. Let $k$ be such that $L_k:=L-2\pi k\in(-\pi,\pi]$. Then equation \eqref{long_eq2} implies that
\begin{equation}
    \de k_0=-i\frac{\sqrt{2}|\spk|\s}{M}\sqrt{-\tmu+\ln(e^{iL_k}-1)}=-i\frac{\sqrt{2}|\spk|\s}{M}\sqrt{-\tmu+\ln(e^{iL}-1)},
\end{equation}
where $\de k_0=k_0-\spk^2/(2 M)$ and the different branches of the logarithm can be taken. For $\tmu\ll-2\pi$, we have approximately
\begin{equation}\label{de_k0_long}
    \de k_0\approx -i\frac{|\spk|\s}{M}\sqrt{-2\tmu+4\pi in}=-i\frac{|\spk|\s}{M}\sqrt{2\ln\frac{\s^3}{\sqrt{2}\pi^{3/2}\rho(x)}},\qquad n\in \mathbb{Z}.
\end{equation}
For the principal branch of the logarithm, the relative magnitude of this correction to the real part of the energy of the longitudinal plasmon-polariton is small for $|\spk|\gg\s$.

Analogously, we have from \eqref{transv_disp_law}, \eqref{Phi_k_FD}, and \eqref{F_D_Im_x_neg} for the transverse modes,
\begin{equation}
    \ln(1+e^{\tmu-x_+^2})=-iR_k,
\end{equation}
where $R_k=R+2\pi k\in[-\pi,\pi)$ and
\begin{equation}
    R:=\frac{|\spk|\s e^{\tmu}}{\sqrt{2\pi}\mu_p^2(x)M}=\frac{\pi |\spk|}{\mu_p^2 M\s^2}.
\end{equation}
As a result,
\begin{equation}\label{de_k0_trans}
    \de k_0=-i\frac{\sqrt{2}|\spk|\s}{M}\sqrt{-\tmu+\ln(e^{-iR}-1)}.
\end{equation}
Thus we see that there are the longitudinal and transverse plasmon-polaritons with the dispersion law close to the dispersion law of plasmons $k_0=\spk^2/(2 M)$ in an equilibrium nondegenerate neutron gas, too. Moreover, there is an infinite number of branches of their dispersion law at a fixed momentum $\spk$. Due to a logarithmic dependence of the corrections \eqref{de_k0_long}, \eqref{de_k0_trans} on the particle density, $\s$, and the plasmon-polariton momentum $|\spk|$, these small corrections agree by the order of magnitude with the analogous corrections \eqref{y_longitud}, \eqref{y_transv} in the case of the Maxwell-Boltzmann distribution. This confirms the robustness of the existence of plasmon-polaritons.

Moreover, it is not difficult to prove that the energies of plasmon-polaritons depend continuously on the parameters of the one-particle density matrix. Consider, for example, the case where the approximations used in Secs. \ref{Long_Plams_Polar}, \ref{Trans_Plasm_Polar} are valid. As is seen from \eqref{int_I}, \eqref{plasm_funk} and \eqref{F_D_func}, \eqref{F_D_Im_x_neg}, the positions of the roots of equations \eqref{disp_law_longit0}, \eqref{disp_law_transv} in the lower complex half-plane of the variable $k_0$ are determined in essence by the distribution function with respect to the longitudinal component of the momentum,
\begin{equation}
    f(p_\parallel):=\int d\spp_\perp \rho(x,\spp),
\end{equation}
regarded as an analytic function of the complex variable $p_\parallel$. Suppose that $f(x)$, $x\equiv p_\parallel\in \mathbb{C}$, depends on the parameter $\la$ and $f(\la,x)$ is a continuous function of the variables $\la$ and $x$. Then the locations of the roots of equations \eqref{disp_law_longit0}, \eqref{disp_law_transv} depend continuously on $\la$. In particular, if $f(0,x)$ is a Gaussian as in \eqref{plasm_funk}, then for a small $\la$ the positions of these roots are close to those found in Secs. \ref{Long_Plams_Polar}, \ref{Trans_Plasm_Polar}. In Secs. \ref{Long_Plams_Polar}, \ref{Trans_Plasm_Polar} and in Appendix \ref{Plasm-Pol_on_Singl_El_App}, we also saw that, in changing the plasmon-polariton momentum, one of these roots can move continuously to the real axis and leave it. All these properties certify that the existence of plasmon-polaritons is not an artifact of the approximations made.

\end{document}